\documentclass[a4paper,fleqn]{cas-sc}
\usepackage{subcaption}
\usepackage[authoryear]{natbib}

\def\tsc#1{\csdef{#1}{\textsc{\lowercase{#1}}\xspace}}
\tsc{WGM}
\tsc{QE}
\tsc{EP}
\tsc{PMS}
\tsc{BEC}
\tsc{DE}

\begin{document}
\let\WriteBookmarks\relax
\def\floatpagepagefraction{1}
\def\textpagefraction{.001}


\shorttitle{Do Persuasive Designs Make Smartphones More Addictive?}
\shortauthors{Chen et al.}

\title [mode = title]{Do Persuasive Designs Make Smartphones More Addictive?  - A Mixed-Methods Study on Chinese University Students}

\author[1]{Xiaowei Chen} [orcid=0000-0003-0794-1551]
\cormark[1]
\ead{xiaowei.chen@uni.lu}
\author[2]{Anders Hedman}
\ead{ahedman@kth.se}
\author[1]{Verena Distler}
\ead{verena.distler@uni.lu}
\author[1]{Vincent Koenig}
\ead{vincent.koenig@uni.lu}

\address[1]{HCI Research Group, University of Luxembourg, Avenue de l'Universit\'e, 4369, Esch-sur-Alzette, Luxembourg.}
\address[2]{Division of Media Technology and Interaction Design, KTH Royal Institute of Technology, SE-10044, Stockholm, Sweden.}
\cortext[cor1]{Corresponding author.}

\begin{abstract}
Persuasive designs become prevalent on smartphones, and an increasing number of users report having problematic smartphone use behaviours. Persuasive designs in smartphones might be accountable for the development and reinforcement of such problematic use. This paper uses a mixed-methods approach to study the relationship between persuasive designs and problematic smartphone use: (1) questionnaires (N=183) to investigate the proportion of participants having multiple problematic smartphone use behaviours and smartphone designs and applications (apps) that they perceived affecting their attitudes and behaviours, and (2) interviews (N=10) to deepen our understanding of users’ observations and evaluations of persuasive designs. 25\% of the participants self-reported having multiple problematic smartphone use behaviours, with short video, social networking, game and learning apps perceived as most attitude and behaviour-affecting. Interviewees identified multiple persuasive designs in most of these apps and stated that persuasive designs prolonged their screen time, reinforced phone-checking habits, and caused distractions. Overall, this study provides evidence to argue that persuasive designs contribute to problematic smartphone use, potentially making smartphones more addictive. We end our study by discussing the ethical implications of persuasive designs that became salient in our study. 

\end{abstract}


\maketitle

\section{Introduction}

Fogg was one of the first scholars who researched computers as persuasive technologies \citep{fogg1998persuasive}. In the early days of persuasive technology, designers focused on promoting positive attitudes and behaviour changes in health, economics and education fields \citep{hamari2014persuasive}. Later, persuasive designs became ubiquitous online, with the adoption of attitude and behaviour change design methods being increasingly integrated into social networking, retail, news and entertainment sites. Billions of users are exposed to such persuasive designs each day by multiple internet giants such as Facebook, TikTok, Amazon, Netflix, Alibaba, YouTube and others.

Persuasive designs might negatively affect users' attitudes and behaviours \citep{borgefalk2019ethics}. On the one hand, for products designed to serve their customers better, there are possibilities that good intentions might cause unintended impacts on the users. One prominent case is the introduction of the Facebook ``like'' button, which was intended to enable users to share affirmation and positivity easily \citep{Rosenstein}. However, studies have shown that the ``like'' button negatively affects users' mental health, resulting in social comparisons and increased envy and depression \citep{blease2015too}. On the other hand, in the context of the attention economy, persuasive designs (for example, recommendation algorithms of video and e-commerce platforms) insatiably seek users' attention and consume their leisure time \citep{williams2018stand}. Companies adopt persuasive designs to prolong user's time on their digital services and seek to make their products more engaging and habit-forming than those of their competitors \citep{eyal2014hooked}. Experts fear that the precipitous rise in smartphone and social media usage \citep{rosenquist2021addictive} is leading to mental \citep{lei2020relationship} and physical harm \citep{kim2015relationship}, especially among children and adolescents \citep{lewis2017our}.

Screen time and mental/physical problems associated with smartphone use are increasing worldwide \citep{olson2022smartphone,busch2021antecedents,rozgonjuk2018association}. Studies found that some mobile applications (\textbf{apps}) predominantly occupy users' screen time, i.e., lifestyle, social networking \citep{noe2019identifying} and instant message apps \citep{ding2016beyond}. Furthermore, No\"e et al. found some features of Snapchat that make users prolong their usage, for example, making friends compete against each other for the top position (competition) and incentivising users not to break their ``Snapstreaks'' with close friends (reward) \citep{noe2019identifying}. It seems legitimate to consider Snapchat's success in light of its integration of these persuasion strategies.

The majority of studies on applying persuasive technology are in health, wellness, and education domains\citep{devincenzi2017persuasive,orji2018persuasive}; recently, scholars have paid growing attention to its adverse effects regarding volunteerism, privacy, transparency, and users' awareness \citep{nystrom2020persuasive}. However, the relationship between persuasive designs and problematic smartphone use (\textbf{PSU}) is less studied. This study makes two main contributions. \\
(1) Our findings extend existing knowledge on how the influence of persuasive design features is perceived by users. \\
(2) We identify potential relationships between persuasive designs and problematic smartphone use behaviours, and links to addiction-like behaviours.  

Our study makes empirical contributions of relevance for researchers and designers in Human-Computer Interaction, as well as psychologists, by providing qualitative insights into the longitudinal effects of persuasive designs on technology users. This study is among the first to use a mixed-methods approach to investigate the relationship between problematic smartphone use and the design features of smartphones in China, one of the most problematic smartphone use countries \citep{busch2021antecedents,olson2022smartphone}. This is timely in an era in which many cutting-edge technologies are applied to exert influence on users across the world.


The remainder of this paper is organised as follows. In section 2, the definitions, applications and ethical concerns of persuasive technology and studies about PSU are examined; subsequently, research questions are formulated. In section 3, the study methods and data analysis process are described in detail. We then present our results in section 4 and discuss the results in section 5. In section 6, we present our conclusions.

\section{Related work}

\textbf{Definitions and applications of persuasive technology:} \cite{fogg2002persuasive} defined persuasive technology as ``\emph{interactive computing systems designed to change people’s attitudes and/or behaviours, without using coercion or deception}''(p.15) (PT1). He excluded unethical applications from the definition. However, when \cite{kampik2018coercion} studied the persuasive properties of several popular apps, including Duolingo, Facebook, Slack, and YouTube, they found prevalent use of deception and coercion and suggested redefining persuasive technology as ``\textit{any information system that proactively affects human behaviour, in or against the interests of its users}'' (PT2). \cite{kampik2018coercion} listed four core indicators of persuasive technology, i.e., intentionally persuasive, behaviour-affecting, technology-enabled, and proactive ("\textit{some extent of autonomy or a high degree of automation}"). This study adopts the PT2 definition and applies the four core indicators in identifying persuasive designs.

Over the past two decades, multiple frameworks and psychological theories have been introduced to the persuasive technology design process as persuasion strategies. \cite{oinas2009persuasive} developed Fogg's taxonomy of persuasive design principles and proposed a framework for the design and evaluation of persuasive systems—the Persuasive System Design (PSD) model (see Figure 1). The PSD model divides persuasion strategies into four categories: primary task support, dialogue support, system credibility support, and social support. \cite{orji2018persuasive} analysed 85 articles on persuasive technologies for health and wellness. They found that the most employed strategies in these cases were ``tracking'', ``monitoring'', ``feedback'', ``social support, sharing and comparison'', ``reminder'', ``alert, reward, points, credits'', ``objectives'', and ``personalisation''. \cite{kaptein2015personalizing} presented persuasion profiling as a method to personalise messages, which represented different persuasion principles, to influence users. \cite{oyibo2019relationship} studied how participants with different personality traits exhibit different susceptibility to social influence strategies (i.e., social learning, social proof and social comparison). All these studies advance our knowledge of how different persuasion strategies could be deployed in designing persuasive applications.

\begin{figure}[ht]
\begin{center}
\includegraphics[width=\linewidth]{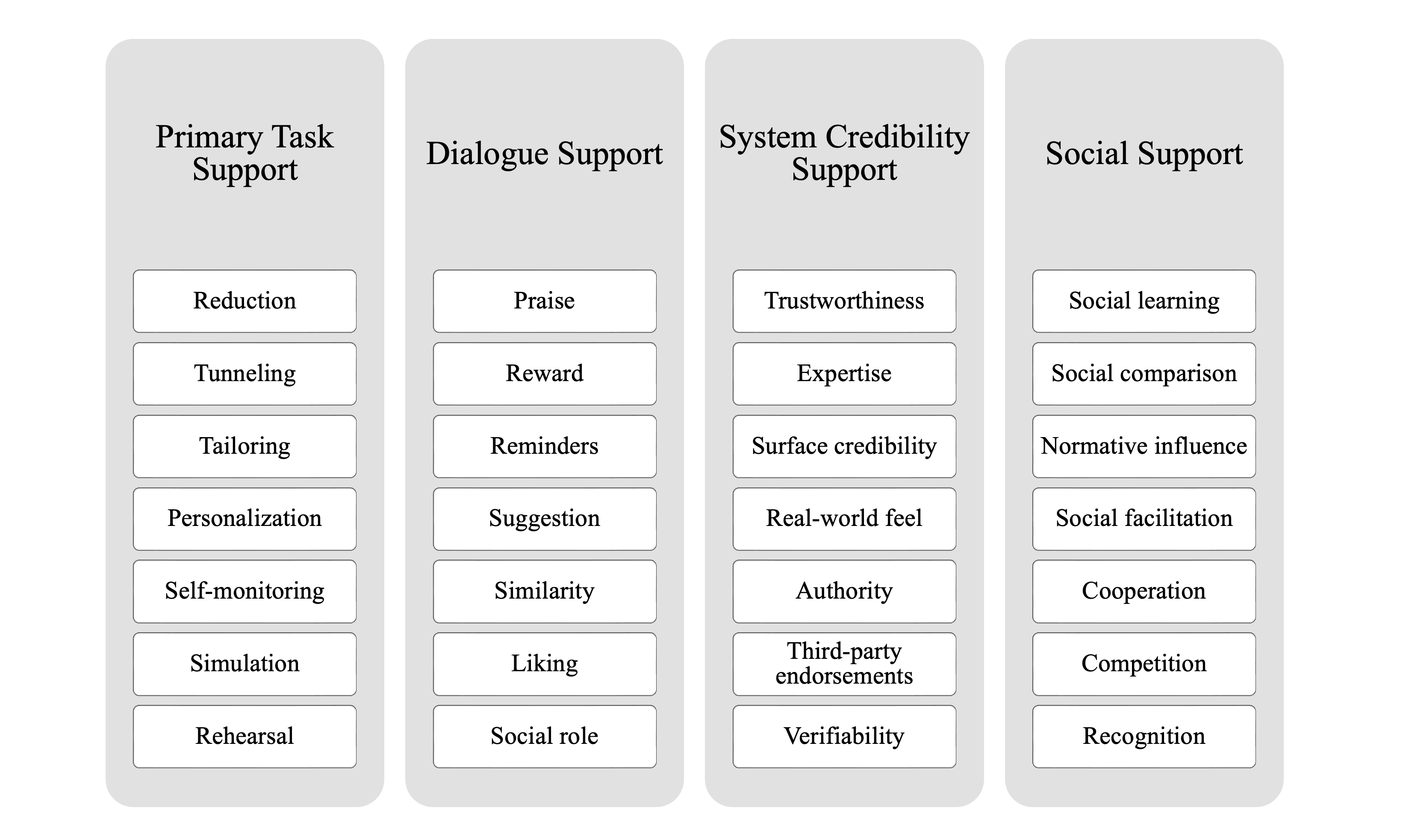}
\end{center}
\caption{The PSD model \citep{oinas2009persuasive}}
\label{fig-1}
\end{figure}

\textbf{Ethical concerns of persuasive technology:} \cite{berdichevsky1999toward} discussed the potential negative impacts of persuasive technology on its users and proposed a guideline for designers. They formulated the golden rule: interaction designers should never seek to persuade users of "\textit{something they themselves would not consent to be persuaded to do}"\citep{berdichevsky1999toward}. \cite{fogg2002persuasive} was concerned about the ethical issues of persuasive technology in the same way as those for persuasion in general and recommended designers perform stakeholder analysis in complicated situations. In addition, Fogg predicted that persuasive technology might encounter increasing scrutiny from policymakers because of its potential impact on the public, prompting regulatory efforts to guard against certain coercive tactics and protect specific audiences. \cite{smids2012voluntariness} found that persuasive technologies often influence users beyond self-control. \cite{smids2012voluntariness} recommended persuasive technology designers perform voluntariness assessments, specifically, reviewing whether there is manipulation ("\textit{influence users in ways of which they are not aware and cannot control}") and coercion ("\textit{apply direct force or credible threat to control users}") in their persuasive designs and whether users act intentionally when being persuaded. \cite{timmer2015ethical} analysed new challenges brought by the integration of persuasive technology and ambient intelligence. They emphasized transparency about the interests of different stakeholders and proposed that users and providers should have open discussions and reach agreements on the goals, methods and interests of persuasive technology to achieve trustworthiness. \cite{borgefalk2019ethics} observed the rise and proliferation of digital platforms that use persuasive designs in business operations, proposing interdisciplinary research approaches, which combine persuasive technology, governance, and management studies, to address associated ethical challenges.

\textbf{PSU and smartphone addiction:} There is a divide in academia on choosing the precise term to describe users' prolonged screen time and mental/physical problems associated with smartphone use \citep{billieux2015can}. Based on the internet addiction scale, \cite{kwon2013development} validated a self-diagnostic scale for smartphone addiction. More than 3300 articles cited the scale and its short version \citep{kwon2013smartphone} as of April 2023. By contrast, some scholars use the term PSU to study the ``\textit{craving to use a smartphone in a way that is difficult to control and leads to impaired daily functioning}'' \citep{busch2021antecedents}. \cite{lanette2018much} noted that the expression ``smartphone addiction'' was overused in academia and the media. They doubted the objectiveness of self-reported scales and stressed the situated and complex nature of phone use. They suggested looking deep into functions and users' self-reflective behaviour. Likewise, \cite{billieux2015we} expressed caution in overpathologizing everyday behaviour as behavioural addictions. They highlighted the multi-faceted nature and heterogeneity of daily life disorders. Therefore, this study does not aim to examine which terms describe current smartphone use problems more precisely; it sets a goal of studying factors that contribute to the problematic situation, and studies using both terms (``smartphone addiction'' and ``PSU'') will be referred to.

\cite{almourad2020defining} reviewed different definitions of digital addiction from 47 studies, including internet, gaming and smartphone addiction. Several features were identified and classified into categories that help to picture digital addiction with respect to four aspects, i.e., device usage, social interaction, psychological states, and clinical symptoms (see Figure 2). In addition to giving a holistic view of the digital addiction research field, they observed a set of shared features across terms, including impulsivity, compulsion, lack of control, negative emotional outcomes, and impairment to work and study \citep{almourad2020defining}. This integrated approach considers the diversity of smartphone functionalities, given that smartphones have become synthetic devices for communication, connectivity, and gaming.

\begin{figure}[ht]
\begin{center}
\includegraphics[width=\linewidth]{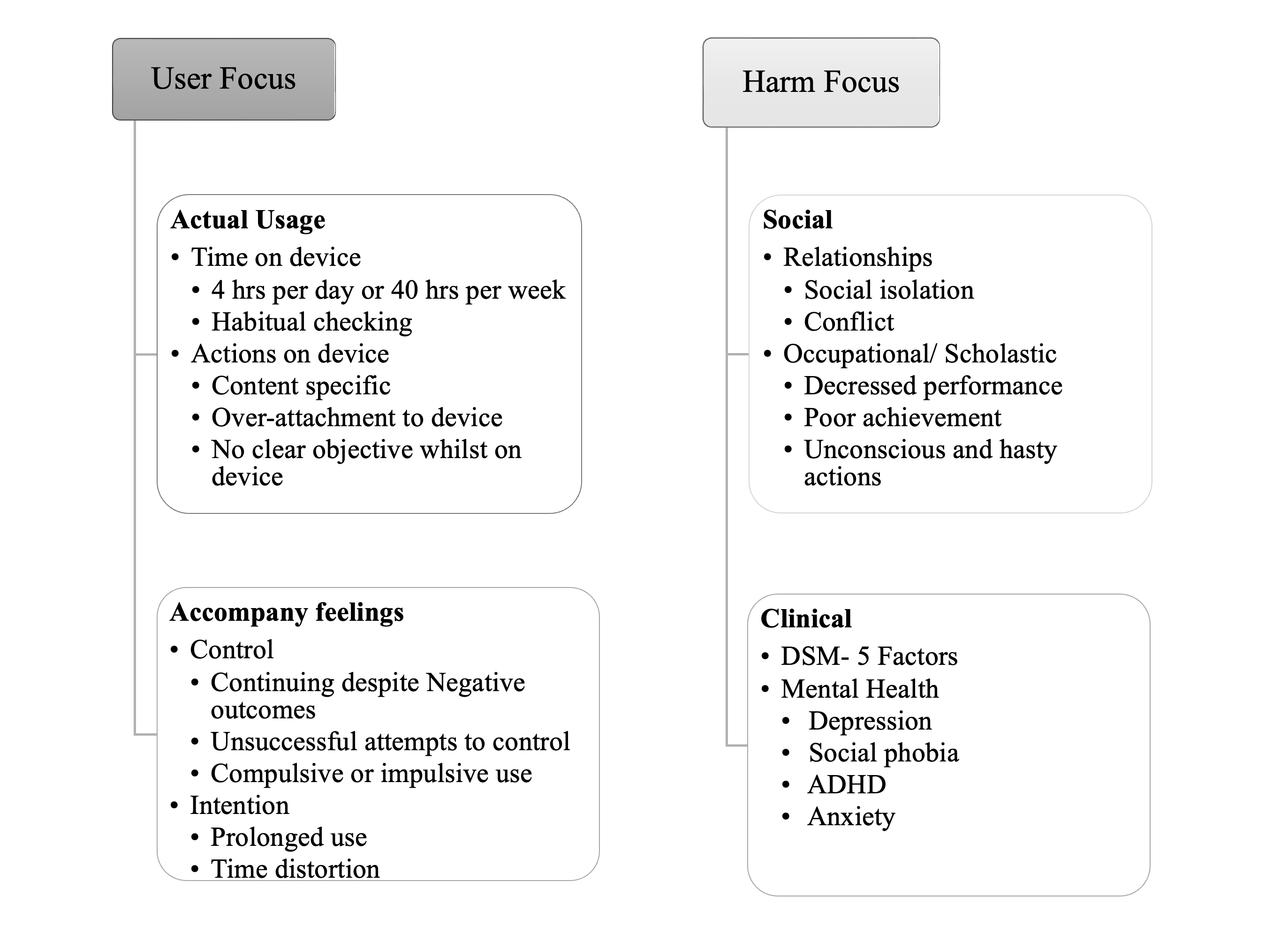}
\end{center}
\caption{Digital Addiction features \citep{almourad2020defining}}
\label{fig-2}
\end{figure}

For the purposes of this article, we will use the term Problematic Smartphone Use (\textbf{PSU}) to refer to users overusing smartphones to the degree which causes a perceived negative impact on their productivity, mental/physical health, and other aspects of life.  

Scholars have striven to seek the factors that contribute to PSU. Some researchers investigated users' susceptibility to PSU. They indicated that user characteristics, such as antagonism and negative affect, could be examined as positive predictors for latent PSU behaviours \citep{marciano2021smartphone}. Further, researchers found that fear of missing out (FOMO)/lower well-being/distress were correlated with PSU \citep{elhai2016fear, horwood2019problematic, della2022psychological}. Higher levels of FOMO were found to be associated with greater impact of social media on one's daily activities and productivity \citep{rozgonjuk2020fear}.Meanwhile, scholars studied users' interactions with smartphones to explore why smartphones are disruptive. Frequent phone-checking habits make up a substantial amount of users' smartphone use \citep{oulasvirta2012habits}. \cite{heitmayer2021smartphones} found that 89\% of smartphone interactions are initiated by users, not by notifications. Importantly, they suggested investigating the routines and habitualised behaviours users developed over time in interacting with smartphones. One possible explanation is that persuasive designs in smartphone operating systems and apps cultivate and reinforce users' habitual behaviours. Studies identified persuasive designs in popular apps such as YouTube, Facebook, Snapchat, Flipkart and AliExpress \citep{kampik2018coercion, noe2019identifying, adib2021systematic}, while persuasive technology is effective in motivating users to change behaviours (empirical studies) \citep{khalil2013harnessing,cellina2021self} and sustaining behaviours (theoretical framework) \citep{kaptein2010persuasion}.

Teenagers spend prolonged time online with social media and games, which might be linked to persuasive designs \citep{daniel2018our}. \cite{cemiloglu2021fine} compared theories to explain digital addiction behaviours with the PSD model, suggesting that certain PSD strategies, such as reduction (simplifying the tasks), rewards (rewarding target behaviours), social comparison (comparing with the others), liking (attractive to its users visually) and personalisation (offering personalised content or services), may trigger and expedite digital addiction in specific contexts. To the best of our knowledge, there is no empirical study on the relationship between persuasive technology and PSU; as a result, we formulate the following research questions (\textbf{RQ}s) to fill this research gap:

\begin{enumerate}
  \item[\textbf{RQ1.}] What proportion of study participants self-report having multiple problematic smartphone use behaviours?  
  \item[\textbf{RQ2.}] Which smartphone designs and apps do participants report to influence their attitude or behaviour? 
  \item[\textbf{RQ3.}] How do participants perceive persuasive designs and their influences on smartphone use?
\end{enumerate}

\section{Research Methods}

\subsection{Overview}
We used a mixed-methods approach to answer our research questions. We administered a questionnaire to answer RQ1 and RQ2 (section 3.2) and interviewed ten study participants to provide additional answers to RQ1 and answer RQ3 (section 3.3).

\subsection{Questionnaire}

\subsubsection{Participants}
Participants were recruited through multiple channels. We published the link to the questionnaire on the university intranet forum (Beijing Institute of Graphic Communication), student group chats (Energy and Sustainability program of Zhejiang University), and Tencent questionnaire service\footnote{\textbf{Tencent questionnaire service:} a free and professional questionnaire design and distribution platform operated by Tencent company, which is the largest social networking company in China.  } (Survey distribution set to college students aged 18 to 26). 248 users completed the survey. With Tencent's automatic spam screen\footnote{\textbf{Tencent automatic spam filter:} the Tencent machine learning algorithm studied respondents’ answering behaviour, number of answers, question types, and other factors to filter spams.} and manual age-grade consistency check (i.e., Chinese students, in general, start their freshman between 17-19 years old; This age-grade consistency check filtered out participants that deviated over two years from the average number), 183 questionnaires were verified as valid.

The participants of our study are Chinese university students. There were 90 male and 93 female participants, ranging from 18 to 26 years old (mean 21.7, SD=1.8). 83\% (n=152) of the participants use Android smartphones, while 17\% (n=31) use iPhones (see Figure 3 the histograms of age, gender, and operating system). The most common study programs in the survey sample were engineering (n=27, 15\%), economics \& management (n=25, 14\%), computer science (n=23, 13\%), and e-commerce \& marketing (n=13, 7\%).

\subsubsection{Procedure}
We pre-tested the questionnaire with four university students. The first two students were from a Swedish university and went through the pretest questionnaire in English. They provided valuable feedback for constructing the smartphone usage and open-ended questions. Then the pretests were held with two Chinese university students with a Chinese translation of the questionnaire. All pretest sessions were conducted via Zoom. We distributed the Chinese translation questionnaire to collect the data. 

 The questionnaire was published on April 3, 2021, and data collection remained open until April 25, 2021. During this period, no quarantine measures were implemented in the cities of our participants. But participants needed to use smartphones to track their contacts and install e-learning apps. Questionnaires could be answered through web pages, QQ, and WeChat. On average, users spent three minutes finishing the survey. Each valid survey participant received ¥3 (US\$0.5) as compensation.

\begin{figure}[ht]
  \centering
  \includegraphics[scale=0.5]{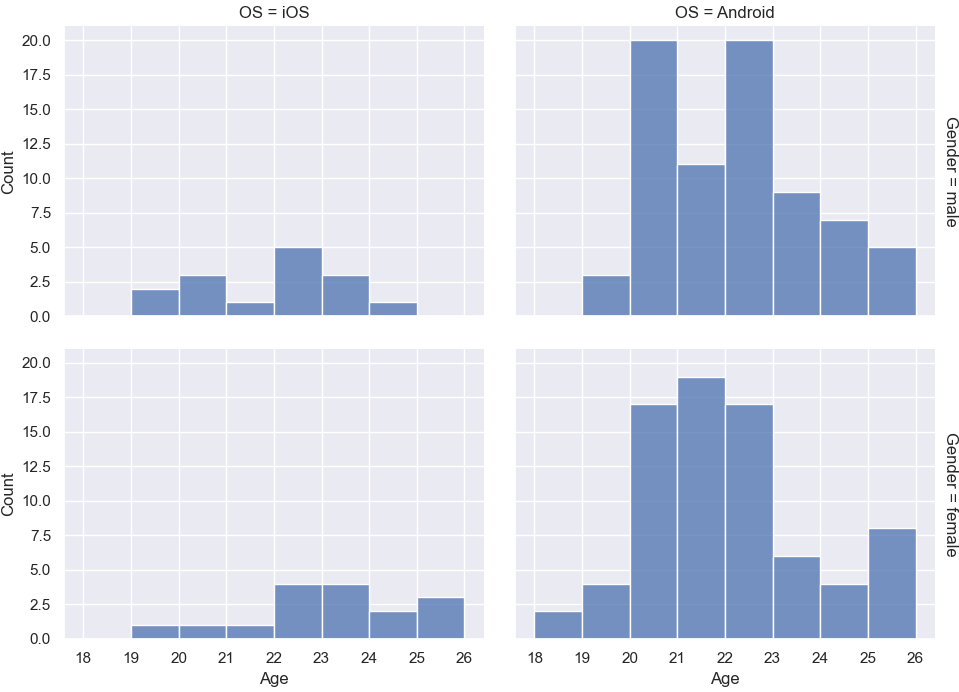}
  \footnotesize{\caption{Age, gender, operating system count}}
  \label{fig-3}
\end{figure}

\subsubsection{Materials}

The questionnaire included multiple-choice questions and open-ended questions in three sections:

\textbf{Demographics} Q1-Q5: Age, gender, smartphone Operating System (OS), study program, and grade (i.e., Bachelor [freshman, sophomore, junior, senior] or graduate program [master's student, PhD student]) of the participants.

\textbf{Smartphone usage} Q6-Q7: we asked the participants their average daily screen time and gaming time. Q8-Q11: We adapted questions from published studies and chose loss of control, perceived negativeness and overuse as key PSU indicators \citep{huang2021network}, including:
\begin{itemize}[ ]
  \item Q8.\emph{ Do you feel that the use of smartphones takes up too much time? \\ a) Yes. \hspace{4mm} b) No. \hspace{4mm} c) Hard to tell. \hspace{4mm}  d) Occasionally.}
  \item Q9.\emph{ Have you tried to control your smartphone usage time? \\a) Yes, I reduced my usage time. \\b) Yes, but I failed to reduce my usage time. \\ c) No, I do not intend to reduce my usage time. \\ d) No, but I plan to reduce my usage time in the future.}
  \item Q10.\emph{ Do you inadvertently use your smartphone for longer times than you planned? \\ a) No. \hspace{4mm} b) Very rarely. \hspace{4mm} c) Rarely. \hspace{4mm}  d) Occasionally. \hspace{4mm}  e) Frequently. }
  \item Q11.\emph{ Does the smartphone negatively affect your studies or professional life? \\ a) No. \hspace{4mm} b) Very rarely. \hspace{4mm} c) Rarely. \hspace{4mm}  d) Occasionally. \hspace{4mm}  e) Frequently. }
\end{itemize}

\textbf{Perception of persuasive applications} Q12-Q13 investigate participants' perceptions of persuasive applications on their smartphones. Specifically, the participants contribute to identifying intentionally persuasive and behaviour-affecting features, and we examine whether the mentioned apps/OS features are technology-enabled and proactive. The questionnaire ends with Q14 about whether the participants would be willing to participate in an interview:
\begin{itemize}[ ]
  \item Q12.\emph{ Are there any apps that changed your attitude or behaviour? (If yes, please elaborate briefly.)}
  \item Q13.\emph{ Are there any functions, apps, or designs of your smartphone that let you to develop new habits? (If yes, please elaborate briefly.)}
  \item Q14.\emph{ Would you like to participate in a 30-minute interview about your smartphone usage habits? (If yes, please leave your contact details.)}
\end{itemize}

\subsection{Interview}

\subsubsection{Participants}
Ten interviewees were sampled from the above survey results with submitted contact information on April 6, 2021; 172 valid answers had been collected by then. Participant gender and screen time were taken into account to match the respective survey sample distribution. We sampled five females and five males. The sampled interviewees reported spending on average 6.0 hrs/d on their smartphones, while the survey participants reported 5.6 hrs/d. Interviewees came from various study programs, including energy and sustainability, computer science, media and civil engineering (see Table 1). 
\begin{table*}
\centering
\caption[caption]{Summary of interviewee information}\vspace{-10pt}
\label{tab-2}
\begin{center}
\begin{tabular}{|c|c|c|c|c|c|c|}
\hline
\textbf{Participants} & \textbf{Ages}  & \textbf{Gender } & \textbf{OS }& \textbf{Study Program} &\textbf{Grade} & \textbf{Screen Time }(hrs per day)  \\ \hline\hline  
P1  &  19   &   Female   &  Androids &  Media  &  Freshman &  6.0  \\ \hline  
P2  &  22   &   Male   &  Androids &  Civil Engineering   &  Senior &  7.0  \\ \hline  
P3  &  22   &   Male   &  Androids &  Computer Science  &  Master Student &  4.0  \\ \hline  
P4  &  22   &   Female   &  iOS &  Mathematics   &  Senior &  6.0  \\ \hline  
P5  &  22   &   Female   &  Androids &  Language  &  Senior &  7.0  \\ \hline  
P6  &  23   &   Female   &  Androids &  Medicine  &  Junior &  8.0  \\ \hline  
P7  &  23   &   Male   &  iOS &  Computer Science  &  Senior  &  6.0  \\ \hline  
P8  &  23   &   Male   &  Androids &  Energy and Sustainability  &  PhD Student &  5.0  \\  \hline  
P9  &  24   &   Male   &  Androids &  Energy and Sustainability  &  Master Student &  5.0  \\  \hline  
P10  &  24   &   Female   &  iOS &  Media  &  Master Student &  6.0  \\ \hline
\end{tabular}
\end{center}
\end{table*}

\subsubsection{Procedure}
We designed and pre-tested interview questions in the same procedure as we described above for designing the questionnaire. The in-depth interviews started with a self-evaluation of matched digital addiction features. Then, we asked interviewees questions related to smartphone usage and screen time. Next, the interviewees were asked to identify persuasive designs that they could observe from their smartphone usage; if the interviewees hadn’t known persuasive technology before the interviews, the definitions, applications and examples of persuasive technology were explained to them. The interviews ended with a self-evaluation of how these persuasive designs observed by interviewees affect their screen time and smartphone usage habits. 

The interviews lasted between 18 and 45 minutes in duration and were conducted remotely via WeChat voice call. The interviews were recorded with permission. Each interviewee was compensated ¥50 (US\$7.3) for participation.

\subsubsection{Materials}

A Chinese translation of Figure 2 was presented to the interviewees, and we provided the definitions of DSM-5 Factors and ADHD as footnotes to the interviewees. We referred to the recommendation and red dots notification designs of WeChat as examples of persuasive technology in question 5, because all the participants were using WeChat and were familiar with these features. We prepared the following seven questions for the interviews:
\begin{enumerate}
  \item \emph{Would you mind going through the Digital Addiction Features (Figure 2) and telling me which features match your experience?}
  \item \emph{Please indicate the occasions when you have to use your smartphone daily.}
  \item \emph{Please evaluate the needed hours for these essential occasions. What are the factors that caused you to spend more time on your smartphone?}
  \item \emph{Have you learned about Persuasive Technology before? (If yes, can you elaborate a bit.)}
  \item \emph{Discuss Persuasive Technology definitions of Fogg (2002) and Kampik et al. (2018) and applications.}
  \item \emph{Can you recognise some persuasive applications/features/designs on your smartphone?}
  \item \emph{Would you mind evaluating the impact of the above-mentioned persuasive applications on your smartphone use?}
\end{enumerate}

\subsection{Data analysis}

The questionnaires were collected in Chinese, and then the raw data were translated from Chinese to English. For screen time and gaming time, mean numbers were computed with respect to gender and operating systems. Quantitative analyses were performed using Microsoft Excel and SPSS. For open-ended questions, sentiment analysis was performed using Excel Azure Machine Learning (ML). The results were also manually checked for errors. Data were visualised using Python Seaborn Library \citep{waskom2021seaborn}.

We used thematic analysis \citep{clarke2015thematic} to analyse the qualitative data collected from interviews. The recorded audios were transcribed to text via the iFlytek\footnote{\textbf{iFlytek:} a Chinese information technology company which is specialized in voice recognition and communication technology. The transcription service in this study was transcribed automatically, and the audio and manuscripts were deleted from the account after the download.} automatic transcription service, then manually checked and extracted by author 1. The interviewees' scripts were translated into English manually by Author 1. Authors 1 and 2 discussed on choosing the matched codes in the coding process and searching for the relevant themes. The themes were reviewed and exchanged between the authors in the process of structuring and composing the results section.

\subsection{Privacy and ethical considerations}

The study was conducted at a Swedish University which does not, under normal circumstances, require an ethical review board for Human-computer Interaction studies. Participants who were younger than 18 years old were automatically deleted from the survey results to protect the privacy of minors. The quantitative and qualitative data collected from the study were analysed anonymously, and the identity of participants was protected. The collected data was stored in an encrypted hard drive. We informed all study participants of the purpose of the study and their right to withdraw. We obtained verbal consent from all the interviewees prior to the interviews.

\section{Results}

\subsection{Questionnaire}
\subsubsection{Problematic Smartphone Usage (RQ1)}
In this section, we describe which proportion of users self-reported having multiple PSU behaviours. 

The participants reported spending on average 5.6 hours/day using their smartphones. 15\% (n=28) of them spent less than four hours per day using their smartphone, while 85\% (n=155) spent four hours or more. On average, female participants used their phones 5.9 hours a day, while male participants used theirs 5.4 hours. We would like to point out that the screen times in this study are self-reported except in Table 3 and interviewees’ scripts which were documented from screenshots. We will explain further in the study limitations.

67\% (n=122) of the participants indicated that they spent too much time on their smartphones. 83\% (n=152) of the participants tried to control their smartphone usage time; among them, 58 (mean, 5.3 hours) participants were able to reduce their screen time, while 94 (mean, 5.8 hours) participants failed to do so. 122 (67\%) participants (frequently and occasionally) used their smartphones for longer times than planned, while 81 (44\%) participants (frequently and occasionally) thought smartphones negatively affected their studies or professional life. 

To sum up, 25\% (n=46) of the participants reported multiple PSU behaviours at the same time, including smartphone overuse (Q8: a or d and Q10: d or e), failure to control or planned to control their phone use in future(Q9: b or d), and felt having been negatively affected by phones (Q11: d or e). They reported spending 6.3 hours daily on their phones.

\subsubsection{Reported persuasive features (RQ2)}
In this section, we describe the perceived persuasive applications reported by survey participants. 

145 (79\%) participants answered the open-ended question: \textit{Are there any apps that changed your attitude or behaviour?} Among the filled-in answers, 38 participants only mentioned app names, with no specification of how these apps influenced them. In consequence of this, 107 valid answers were analysed by Azure to identify the sentiments. The most mentioned apps were TikTok \footnote{\textbf{TikTok:} TikTok in this study refers to Douyin which is the Chinese version of TikTok international version.}, WeChat (social networking), Honor of Kings (a popular game), Kuaishou (short video platform), Little Red Book (social networking), Weibo (Chinese Twitter) and Taobao (e-commerce) (see Table 2, the most mentioned apps that changed users' attitude or behaviour with sentiment analysis, count, and coded comments). These are all popular apps among young Chinese. Surprisingly, TikTok, WeChat, Honor of Kings, and Taobao were most frequently mentioned as having negative influences on the participants.

\begin{table*}
\centering
\caption[caption]{The most mentioned apps that changed attitudes or behaviours of the participants}\vspace{-10pt}
\label{tab-1}
\begin{center}
\begin{tabular}{|c|c|c|c|c|c|c|}
\hline
\multirow{2}{*}{\textbf{Apps}} & \multicolumn{4}{c|}{\textbf{Sentiment Analysis}} & \multirow{2}{*}{\textbf{Count}} & \multirow{2}{*}{\textbf{Comment}}  \\   \cline{2-5} \
& \textbf{Pos}.  & \textbf{Neu}. & \textbf{Neg}. &  \textbf{N.D.} &  &   \\   \hline\hline 
\multirow{3}{*}{TikTok} & \multirow{3}{*}{11} & \multirow{3}{*}{2} & \multirow{3}{*}{10} & \multirow{3}{*}{5} & \multirow{3}{*}{28}  & \multicolumn{1}{c|}{\text{Learn skills/Broad horizon/Inspirational/}}  \\
&  &  &  &   &  &  Kill boredom/New Knowledge/Lose control/Can't stop/ \\
&  &  &  &   &  &  Too much time/You can find everything on it \\   \hline 
\multirow{4}{*}{WeChat} & \multirow{4}{*}{5} & \multirow{4}{*}{2} & \multirow{4}{*}{8} & \multirow{4}{*}{4} & \multirow{4}{*}{19}  & \multicolumn{1}{c|}{\text{Help me with social/Inspiring tweets/Convenient }}  \\
&  &  &  &   &  &  to make payment/Without notice, half of my \\
&  &  &  &   &  &  screen time was on social media/FOMO/\\
&  &  &  &   &  &  Constantly check my phone/Distraction \\   \hline 
\multirow{2}{*}{Honor of Kings} & \multirow{2}{*}{} & \multirow{2}{*}{} & \multirow{2}{*}{3} & \multirow{2}{*}{7} & \multirow{2}{*}{10}  & \multicolumn{1}{c|}{\text{Angry/stay up late/}}  \\
&  &  &  &   &  & Decreased self-control and study time \\   \hline 
Kuaishou & \multirow{2}{*}{4} & \multirow{2}{*}{} & \multirow{2}{*}{3} & \multirow{2}{*}{2} & \multirow{2}{*}{9}  & \multicolumn{1}{c|}{Life tips/Diverse/You can watch everything/Vulgar/}  \\
short video &  &  &  &   &  &  Don't want to put down your phone/not so optimistic\\   \hline 
\multirow{2}{*}{Little Red Book} & \multirow{2}{*}{7} & \multirow{2}{*}{} & \multirow{2}{*}{} & \multirow{2}{*}{2} & \multirow{2}{*}{9}  & \multicolumn{1}{c|}{\text{Learn new knowledge/Different people and diverse world/}}  \\
&  &  &  &   &  &  Help me with dress code\\   \hline 
Learning apps (2); & \multirow{4}{*}{6} & \multirow{2}{*}{} & \multirow{4}{*}{} & \multirow{4}{*}{2} & \multirow{4}{*}{8}  & \multicolumn{1}{c|}{Learn a lot in extracurricular/TED videos/ }  \\
XueXiaoyi (1); &  &  &  &   &  &  Help me with my study/ \\
Xuexi Tong (1);&  &  &  &   &  &  Learn knowledge and ideas \\
Zhihu (4)&  &  &  &   &  & about life and science \\ \hline 
Forest (4); & \multirow{3}{*}{6} & \multirow{3}{*}{} & \multirow{3}{*}{} & \multirow{3}{*}{1} & \multirow{3}{*}{7}  & \multicolumn{1}{c|}{Focus/Reduce time playing my phone/ }  \\
Toma Todo (3)&  &  &  &   &  &  More concentrate on study/  \\
&  &  &  &   &  &  Schedule my study time   \\   \hline 
\multirow{2}{*}{Weibo} & \multirow{2}{*}{4} & \multirow{2}{*}{1} & \multirow{2}{*}{2} & \multirow{2}{*}{} & \multirow{2}{*}{7}  & \multicolumn{1}{c|}{Diverse world/Meet new people/}  \\
&  &  &  &   &  &  Check phone too frequently \\   \hline 
\multirow{2}{*}{Taobao} & \multirow{2}{*}{} & \multirow{2}{*}{2} & \multirow{2}{*}{3} & \multirow{2}{*}{1} & \multirow{2}{*}{6}  & \multicolumn{1}{c|}{Addiction/No longer enjoy shopping malls/}  \\
&  &  &  &   &  &  Spent too much money on it \\   \hline
\multicolumn{7}{|l|}{\small Notes:  Pos. for Positive; Neu. for Neutral; Neg. for Negative; N.D. for No Detail.} \\
\multicolumn{7}{|l|}{\small \qquad\quad `(n)' in Apps means count of the specific app.}\\   \hline
\end{tabular}
\end{center}
\end{table*}

``Time'' was the most frequently mentioned keyword which was recorded in 23 answers. Positive sentiments were associated with Countdown (a timer app with schedule features), Forest (time management), Douban (an online community of book, music and movie lovers), e-learning, Screen Time (iOS and Android digital health functions), and Toma Todo (a timer app with screen locker function). In contrast, negative sentiments were linked to Honor of Kings (decreased self-control), TikTok (spent too much time on TikTok, ``cannot stop''), WeChat (without notice, half of the screen time were spent).

More than ten users positively commented on a set of lifestyle, hobby and learning apps: Keep (fitness), Mint (healthy diet), National Karaoke (hobby), Kuwo Music, Duolingo (language learning), and Fluently Speaking (English learning). Words such as ``fun'', ``inspiring'', ``helpful'', ``time well-spent'', and ``productive'' were found in these positive comments.

125 (68\%) participants answered the question: \textit{Are there any functions, apps, or designs of your smartphone that let you to develop new habits?} Among the filled-in answers, 26 participants did not elaborate on their answers. Azure marked 4 of the answers as negative, 25 as neutral, and 70 as positive. 31 (17\%) participants mentioned the functions of their smartphone operating systems with positive sentiments, such as: ``\textit{AI assistant is so smart, I get used to operating my phone using voice}'', ``\textit{the Digital Health function gives me a clear idea about how much time I spend on my phone}'', ``\textit{I use phone memos to write lab notes, it is so convenient}'', and ``\textit{Turning on NFC by double-clicking makes the payment process easier, saving commute time}''. It was observed that these participants were satisfied with the utilisation and application of forefront technologies, and they accepted and appreciated the convenience brought by smartphones. Many identified useful operating system functions that had nothing to do with persuasive technology.

The most mentioned apps that lead to new habits were WeChat (11 times, about changing ways of socialising and making payments, refreshing app to check updates, walking 10000 steps daily, etc.); Toma Todo (6 times, about concentrating on learning); Alipay (4 times, about digital payment and feeding pets on virtual farms); Baidu (3 times, about the benefits of map and search engine). All these comments were either positive or neutral, except one participant mentioned that ``\textit{WeChat has negative influences on my sleep time}''. Survey participants mentioned a set of apps that led them to adopt daily check-in habits: QQ (instant messaging), Taobao (e-commerce), Banking (finance), Alipay (finance), Forest (time management), vocabulary app and Xuexi Tong (e-learning).

Overall, short video, social networking, game, e-learning and time management apps were the most persuasive apps reported by participants. Habit changes brought by built-in features of smartphones were almost entirely described to be positive, while the impact of apps on users was both positive and negative; Social networking, time management, and digital payment apps were the most mentioned apps that lead to new habits; additionally, some of these apps lead users to daily check-in habits. 

\subsection{Interviews}

\subsubsection{Self-evaluation of smartphone use (RQ1)} 
In this section, we document interviewees' self-evaluations of smartphone use and their reflections. 

 After going through the features of digital addiction (Figure 2), interviewees indicated which experiences matched their own. The most frequently mentioned were using smartphones ``over four hours per day'', ``habitual checking/unconsciously unlocking phone'', ``checking specific content on smartphones'', ``time distortion/forget about time'', and ``prolonged usage''.

P2, P8 and P9 self-reported that they felt addicted to their phones. They reported spending on average 5.7 hours per day on their phones. They mentioned the following symptoms: Time distortion (P2, P8), mild depression (P9), low productivity in studies (P8, P9), blurring eyesight and sore fingers (P2), and habits that could cause physical harm (P8 and P9 share a common habit of scrolling phones while walking). For participants who do not consider themselves addicted, some symptoms were self-evaluated: bad sleep quality associated with phone usage (P10), sometimes feeling anxious when scrolling the phone (P1), cannot stop scrolling (P4, P6) and habitually unlocking phones (all participants except P7).

P8 and P10 expressed that they were worried about spending over two hours daily on WeChat to socialise with peers because of FOMO. P2, P3 and P5 expressed that their performance in study/internship had been less productive recently due to excessive use of smartphones:

``\emph{I know that I spend too much time on my smartphone, it negatively affects me. I cannot focus on studying and often drift away. Tried a few times to reduce screen time; however, I never succeeded.}'' (P2)

``\emph{Playing with my smartphone causes me to delay the hand-in of assignments. When the stress is high, it is more difficult to put aside my phone. This leads to a cycle of inefficiency and self-indulgence.}'' (P3)

``\emph{I was troubled by the notifications. I fear that I will miss something important if I do not read them. Some make me emotionally disturbed, which affect my study and productivity.}'' (P5)

In order to gain a deeper understanding of the roles of smartphones in the interviewees' daily lives, the interviewer asked in which situations the interviewees must use their smartphones and with what functions. According to these functionalities, a list of apps could then be identified and divided into six categories:
\begin{enumerate}[ ]
  \item \textbf{Social networking} QQ, WeChat, Weibo, Douban, Little Red Book;
  \item \textbf{Shopping} Taobao, Pinduoduo, JD, Alipay, Xianyu, Meituan takeaway;
  \item \textbf{Study/work} DingTalk, university apps, Email, NFC commute card;
  \item \textbf{Tools}  Vocabulary apps, maps, Forest, stocks and funds, banks, Toma Todo, Calendar;
  \item \textbf{Reading} Zhihu, WeChat news subscription, Qidian online;
  \item \textbf{Leisure} Music apps, games, short video apps, streaming services.
\end{enumerate}

When the interviewees were asked to evaluate the needed hours for their essential use, the hours ranged from 1 hour to 5 hours. The mean time was 3.5 hours, which is 58\% of their self-reported screen time (mean, 6 hours). The interviewees mentioned some of the factors that caused them to spend prolonged hours:

``\emph{I feel bored when commuting, so I play with my smartphone like others; When I encounter difficulty in writing my bachelor thesis and my internship tasks, I check social media and escape from all the stress during the break to relax.}'' (P5)

``\emph{My roommate and I compete on the Alipay virtual farm. It is a silly game; however, it is fun to have a routine game with a friend. Additionally, I play TikTok videos when I have meals; then, time flies without notice.}'' (P6)

``\emph{I know that using a smartphone for 6 hours daily is a bit too much. The entertainment provided by the smartphone is very convenient. Since I am so happy when playing on the device, and making changes will be painful, why do I need to control my usage? If the purpose of life is to pursue happiness, smartphones can indeed fulfil my needs.}'' (P7)

``\emph{When I hang out with my friends during the weekends, I have less screen time. However, when I spend weekends by myself, I feel isolated if I don’t refresh my social media feeds. Also, commuting between two campuses of the university takes 3 to 4 hours each week. I play with my phone on public transportation.}'' (P9)

Participants used phrases associated with emotions in their own reflection on overusing smartphones: ``bored'', ``stress'', ``relax'', ``fun'', ``happy'', ``isolated''. 

In summary, three out of ten interviewees self-reported that they were addicted to their smartphones. Physical harm, depression, FOMO and low productivity were reported that associated with smartphone overuse. Interviewees reported apps that were essential to them, covering many aspects of their life. Interviewees used emotion-related expressions to reflect on their smartphone overuse.

\subsubsection{Perception of persuasive designs (RQ3)}
In this section, we describe how interviewees perceived persuasive designs and which persuasive designs interviewees identified from their smartphones.

Interviewees were divided on knowledge of persuasive technology. On the one hand, two interviewees from computer science (P3 and P7) and one interviewee from a media major (P10) had learnt about persuasive technology in their previous education. They recalled persuasive technology as ``\textit{using automatic algorithms and notifications to persuade users}'' (P3), ``\textit{applying psychological methods in ICT products design}'' (P7), and ``\textit{studying users' preference to cultivate and reinforce habits}'' (P10). On the other hand, the other seven interviewees had no prior knowledge of persuasive technology before the interviews. To investigate their perceptions of persuasive technology, definitions, applications and examples of persuasive technology were discussed with interviewees to ensure that the interviewees understood what persuasive technologies were and how they worked. After the discussion, the interviewer invited the interviewees to identify persuasive applications and features in the apps they use daily. The persuasive apps, designs and features that the interviewees identified can be classified into the following categories:

\textbf{Social networking} Little Red Book integrates purchase links into its online community, making it easier to place orders from the influencers' posts. WeChat subscriptions, QQ notifications, Weibo and Douban recommend articles and ads based on the users' viewing history and profiles. WeChat uses tags, such as ``N friend(s) favourited'' and ``N friend(s) read this article'' to persuade users to click the recommended articles (see Appendix Figure 4(b). All these social networking companies incorporated ``like'' buttons in their apps.

\textbf{Shopping} Taobao, Pinduoduo, Xianyu, and JD recommend new purchases based on users' searching, browsing and typing history. Meituan takeaway suggests menus to users according to their location, profile, and location weather forecast. Pinduoduo uses \emph{\textbf{gambling-like designs}}\footnote{\textbf{Gambling-like designs:} Pinduoduo persuades users to invite their friends to download or register with Pinduoduo to endorse them to get ``free cash'' or ``free products''. Pinduoduo's algorithm calculates how much the invited friends would endorse a user. In many cases, even if the user spends days inviting all his friends to sign up and log in to Pinduoduo, the promised money has never been transferred to the user. This persuasion strategy applies deception and matches the definition of gambling, i.e., ``the practice of risking money or other stakes in a game or bet'' (merriam-webster.com). In these gambling-like designs, users lost their valuable time and friends' personal data; and most users gained nothing at the end of the games. Lawyer Liu Yuhang sued Pinduoduo for allegedly violating the principle of good faith, using false data, and concealing rules, which constituted fraud, on March 31 2021, in Shanghai, China. Even though Pinduoduo lists on Nasdaq, there is not much international coverage of this lawsuit which has been trending on Chinese social media sites for weeks in April 2021.} to deceive users and gamification to attract users to spend more time on it (see Appendix Figure 4(a)). Alipay's virtual farm uses incentives (virtual coins) and competition (ranking) to foster users' daily check-in habits. Additionally, all these apps send coupons to stimulate new purchases.

\textbf{Tools} Time management apps such as Toma Todo and Forest have persuasive reminders to help users manage their schedules. Keep and Mint have persuasive notifications to encourage users to exercise and eat healthy diets. Vocabulary apps use personalised notifications and goals to remind users to check in daily. Baidu map highlights restaurants and shops that paid promotion fees to make their locations more visible.

\textbf{Reading} Top Buzz news and Zhihu recommend articles based on users' reading history. Ads were personalised according to users' unique profiles, making the ads more attractive and relevant to users. Top Buzz, WeChat subscription channel and Weibo trending news use emotionally triggering titles to lure users. Machine learning (ML) algorithms were applied to provide personalised suggestions.

\textbf{Leisure} Short videos, user-generated content, and streaming platforms recommend new videos/playlists based on users' viewing history (i.e., WeChat video, TikTok, Kuaishou, iQIYI, Youku, Bilibili, YouTube). TikTok and Kuaishou integrate buying links with video content, encouraging users to place orders with only one click.

\textbf{OS} The red dots on app and system icons draw users' attention and keep persuading users to click on them (iOS and Android). Xiaomi and Huawei phones show recommendations of readings and app promotions after system updates. Participants were annoyed by those solicitations, which were difficult/impossible to turn off.

Proactive persuasive designs were identified in most essential apps by interviewees, except for a few tools (stocks \& investments, calendar) and study/work apps (DingTalk, university, email and NFC commute apps). Interviewees used negative expressions to describe their experience with intrusive persuasions that relied on personal information and distracting notifications. On the other hand, interviewees used positive expressions to discuss persuasive features of time management apps, Keep, Mint, and vocabulary apps. The interviewees used neutral expressions to describe their experience with persuasive designs, but they were annoyed that they could not turn off some persuasive features that were imposed on them without consent.

All interviewees expressed positive sentiments towards their phones. This matches the analysis of the questionnaire’s fill-in answers. However, users complained about the red dots on the operating system icons, which made users click reflexively and intrusive AI recommendations (for example, the AI assistant page of Harmony OS), which were described as “harassing” and "manipulative" by participants. E-commerce apps recommend users too many products according to their purchase history and browser history. P3, P7, and P10 identified more persuasive designs than other interviewees. Concerns about user privacy and data ownership were also raised by P7 and P10.

In summary, interviewees identified multiple persuasive designs in most of their essential daily  apps, and they expressed quite positive sentiments towards the functionalities of these apps; however, they complained about the distraction, lack of consent, user privacy and blurring data ownership related to persuasive designs. Besides the common persuasion strategies, some ethically controversial ones were reported by interviewees, that is, using emotional triggers and gambling-like designs to persuade users.

\subsubsection{Evaluation of persuasive designs (RQ3)}
In this section, we record how the interviewees evaluate the influences of persuasive designs on them. 

Interviewees indicated that persuasive designs prolonged their screen time, caused distractions, and affected their behaviours; specifically, P5, P7 and P10 mentioned the disruptiveness, exploiting human weakness and reinforcing habits of persuasive designs:

``\emph{Without these subtle persuasive designs, I would spend less time checking my phone. When I was reading a paper or working on assignments, the reminders of my followed drama just updated would prompt me to watch the drama first.}'' (P5)

``\emph{All popular apps have persuasive designs integrated. Interaction designers took advantage of human weaknesses and created these apps. It seems that only users are being blamed for their lack of self-control.}'' (P7)

``\emph{Persuasive designs feed users according to their preferences, which could make users stuck in their own habits. Knowing how it works does not make me immune to the algorithms.}'' (P10)

At the end of the interviews, interviewees were asked to spend two minutes reflecting on their smartphone usage and evaluating the influence of persuasive apps on them. All interviewees indicated that persuasive applications increased their smartphone usage time. The interviewees estimated that if they could turn off all persuasive features on their smartphones, they might reduce screen time by 10\% to 65\%, with a mean value of reducing 37\% (see Table 3).

\begin{table*}
\normalsize
\centering
\caption[caption]{Screen time and App usage of interviewees}\vspace{-10pt}
\label{tab-3}
\begin{center}
\resizebox{\columnwidth}{!}{
\begin{tabular}{|c|c|c|c|c|c|}
\hline
\multirow{2}{*}{\textbf{Participants}} & \multicolumn{1}{c|}{\textbf{Screen Time}} & \multicolumn{1}{c|}{\textbf{Necessary}} & \multirow{2}{*}{\textbf{PD}} & \multirow{2}{*}{\textbf{Comments}} & \multicolumn{1}{c|}{\textbf{Most used apps}} \\ 
&  (hrs per day)  &  (hrs per day) & & & (hrs per day)  \\   \hline\hline 
\multirow{2}{*}{P1} & \multirow{2}{*}{7.8} & \multirow{2}{*}{3.0} & \multirow{2}{*}{-30\%} & \multicolumn{1}{c|}{No intention to reduce smartphone usage.} & \multicolumn{1}{c|}{Weibo (2.0); WeChat (1.2); } \\
&    &   & & Smartphone is a useful tool.  &  TikTok (1.1); QQ (0.7)  \\   \hline 
\multirow{2}{*}{P2} & \multirow{2}{*}{10.0} & \multirow{2}{*}{2.5} & \multirow{2}{*}{-50\%} & \multicolumn{1}{c|}{I blocked all notifications and try to reduce} & \multicolumn{1}{c|}{WeChat (2.4); Videos (1.4);} \\
&    &   & & the frequency of checking my phone.  &  TikTok (1.4); Gaming (1.1)  \\   \hline 
\multirow{2}{*}{P3} & \multirow{2}{*}{4.0} & \multirow{2}{*}{2.0} & \multirow{2}{*}{-50\%} & \multicolumn{1}{c|}{I blocked notification, recommendation,} & \multicolumn{1}{c|}{WeChat (1.6); Browser (0.7);} \\
&    &   & & and deleted video apps to reduce my screen time.  &  Music (0.5); Stock app (0.3)  \\   \hline 
\multirow{2}{*}{P4} & \multirow{2}{*}{5.7} & \multirow{2}{*}{1.0} & \multirow{2}{*}{-50\%} & \multirow{2}{*}{$\backslash$} & \multicolumn{1}{c|}{Douban (1.6); WeChat (1.2);} \\
&    &   & &   &  YouTube (0.8); TikTok (0.6)  \\   \hline 
\multirow{2}{*}{P5} & \multirow{2}{*}{7.0} & \multirow{2}{*}{3.0} & \multirow{2}{*}{-20\%} & \multicolumn{1}{c|}{I traveled to another city with friends and} & \multicolumn{1}{c|}{Gaming (1.7); WeChat (1.5);} \\
&    &   & & do not have much time to play with smartphone.  &  Meituan (1.4); Weibo (0.4)  \\   \hline 
\multirow{2}{*}{P6} & \multirow{2}{*}{9.1} & \multirow{2}{*}{6.0} & \multirow{2}{*}{-25\%} & \multicolumn{1}{c|}{Smartphone is my only digital device} & \multicolumn{1}{c|}{Videos (1.2); Pinduoduo (1.1);} \\
&    &   & & for assignments, lectures and entertainment.  &  Kuaishou (1.0); WeChat (0.8)  \\   \hline 
\multirow{2}{*}{P7} & \multirow{2}{*}{6.1} & \multirow{2}{*}{5.0} & \multirow{2}{*}{-25\%} & \multicolumn{1}{c|}{I used iOS screen time to} & \multicolumn{1}{c|}{WeChat (2.4); Bilibili (1.5);} \\
&    &   & & control the usage of certain app.  &  QQ Music (0.3); Taobao (0.3)  \\   \hline 
\multirow{3}{*}{P8} & \multirow{3}{*}{8.0} & \multirow{3}{*}{5.0} & \multirow{3}{*}{-65\%} & \multicolumn{1}{c|}{I used multiple devices to reduce my frequency} & \multicolumn{1}{c|}{WeChat (2.8); Bilibili (1.1);} \\
&    &   & &  of checking phone. All work-related tasks &  Ciwei reading (0.5) \\
&    &   & &  have been moved to laptop and iPad. &  Xianyu (0.4) \\   \hline 
\multirow{2}{*}{P9} & \multirow{2}{*}{7.0} & \multirow{2}{*}{2.5} & \multirow{2}{*}{-10\%} & \multirow{2}{*}{I deleted TikTok.} & \multicolumn{1}{c|}{Reading (2.6); WeChat (2.5);} \\
&    &   & &   &  Bilibili (0.8); QQ (0.7)  \\   \hline 
\multirow{3}{*}{P10} & \multirow{3}{*}{6.0} & \multirow{3}{*}{5.0} & \multirow{3}{*}{-40\%} & \multicolumn{1}{c|}{Smartphone has been integrated into} & \multicolumn{1}{c|}{WeChat (1.4);} \\
&    &   & & my daily life, make tasks easier. &  Gaming (1.2); TikTok (1.0);   \\
&    &   & & Smartphone reduced my time of using laptops. &  Weibo (0.7)  \\  \hline 
Mean &7.1 &3.5	&-37\% &-	&-   \\   \hline
\multicolumn{6}{|l|}{\small Notes: screen times and most used apps in this table were logged from screenshots provided by the interviewees.} \\
\multicolumn{6}{|l|}{\small \qquad\quad PD: Persuasive Designs. This column is the user's self-evaluation; if they could turn off all persuasive features, their screen time might change n\%.}\\
\multicolumn{6}{|l|}{\small \qquad\quad `$\backslash$': No record.  `-': not applicable.}\\   \hline
\end{tabular}
}
\end{center}
\end{table*}

All interviewees shared their screenshots of screen time with the interviewer during the interviews. The most frequently used apps by interviewees were social networking apps (WeChat, Weibo, QQ), video platforms (Tencent, Youku, Bilibili), short video apps (TikTok, Kuaishou), shopping apps (Taobao, Pinduoduo, Xianyu, Meituan), reading apps (Ciwei, Qidian) and games. These apps accounted for more than half of the screen time of the interviewees. As we recorded in Section 4.2.2, nearly all these apps integrate multiple persuasive designs into their services. We collected comments from the interviewees three weeks after the interviews regarding their participation in the study. Five interviewees have reduced their screen time after the interviews, and they mentioned blocking notifications, deleting addictive apps, travelling with friends, setting time limits on addictive apps, and using alternative devices to reduce the frequency of unlocking their smartphones. For interviewees who did not seem to worry about their screen time, their screen times remained nearly the same, and they mentioned that smartphone was ``useful'', ``integrated with everyday life'' and one participant's ``only digital device''.

Overall, interviewees evaluated that persuasive designs prolonged their screen time, reinforced habits and caused distractions. Interviewees had different evaluations of the impact of persuasive designs on their screen time, but all reported that persuasive designs increase their screen times.
 
\section{Discussion}

In this section, we reflect on the three research questions of this paper. First, we discuss our results regarding problematic smartphone usage (RQ1). Second, we discuss the results of perceptions, identification and evaluation of persuasive designs (RQ2 and RQ3). Next, we discuss our findings regarding interaction design ethics. We also discuss the limitations of the present work and suggest directions for future research. 

\subsection{Problematic smartphone use: proportions, screen times and reflections}

Throughout our study, similar proportions of participants reported multiple PSU behaviours (25\% in the survey, 3 out of 10 in the interviews). Many previous studies have used SAS-SV to measure the PSU of Chinese students, with varying results, ranging from 30\% \citep{chen2017gender} to 72\% \citep{yang2019chinese}. Different sample methods and participant demographics might contribute to the difference in results. The reported symptoms of interviewees, such as time distortion, mild depression, low productivity, blurring eyesight, and wrist pains, were also recorded by other PSU studies \citep{busch2021antecedents,rozgonjuk2020associations}. 

There are multiple reasons why survey participants reported, on average, spending 5.6 hours a day on their phones. Previous studies found that correlation points between FOMO and PSU are much higher for Chinese students compared to western country studies \citep{elhai2020depression}. The anxiety of COVID-19 pandemic was found to be associated with PSU severity \citep{elhai2020covid}. Besides these reasons reported in earlier work, we found that the less developed IT infrastructures of Chinese universities (intranet and emails), which resulted in the adoption of instant messages apps (QQ and WeChat) as primary communication tools among students, administrators and educators, make students spend prolonged hours on their phones.  


Social network and short video apps are the most mentioned apps in the questionnaire, occupying much of the interviewees' screen time. Users tend to conflate smartphone usage and application usage when self-reporting PSU \citep{rozgonjuk2020associations}. When investigating PSU, researchers should be aware of this problem and take into account gathering information on both general smartphone usage and application-specific usage to better understand the relationship between them. Additionally, different social network applications display varying degrees of addictive potential and impact on users \citep{rozgonjuk2020associations}, underscoring the need to analyse the platforms that users choose and their potential contribution to problematic behaviours \citep{rozgonjuk2021comparing}. 

Interviewees' reflections on necessary usage and overuse revealed that smartphones have become indispensable for study/work, social, leisure and finance. Societies are undergoing large-scale digital transformations globally, moving both public service and private business online. It is almost impossible to live everyday life without smartphones. People would face social, study, mobility, and work difficulties without apps (see Section 4.2.1). The categories of necessary apps reported by interviewees can explain why 85\% of the survey participants spent at least four hours daily with their smartphones. People often cannot refuse to use smartphones when living in digital societies. In addition, we found that participants regard their phones as interactive narrative relays between them and others (online strangers/peers/friends/family members). Smartphone users project their mental life through smartphones and receive the projections of others. These internal and external mental projections interact with each other and have impacts on the user's daily life. Some kinds of companionship are formed between people and their phones, as evidenced by the fact that many users use emotional words to describe their overuse of smartphones. 

\subsection{Persuasive designs are making smartphones more addictive}

We found persuasive designs to be more prevalent than we had expected. Persuasive designs were identified in most daily apps by the interviewees, except for a few essential apps (see Section 4.2.2). Reviewing the apps that changed attitudes or behaviours from the questionnaire results (Table 2), we found that all but a few learning apps were integrated with multiple persuasive designs. We can deduce from these findings that study participants live their everyday life with ambient persuasions. 

Persuasive designs are often designed to exploit users for indirect monetisation by manipulating them to spend as much time as possible interacting with apps and services. For shopping apps, the most identified persuasive designs were personalisation, reduction, suggestion and rewards; for social networking apps, they were recognition, personalisation, social comparison, and reminders; for leisure and readings, the most identified were liking, suggestion, tracking, reduction, and monitoring. These most identified strategies overlapped with Orji and Moffatt's analysis \citep{orji2018persuasive}. However, they found these most employed designs by analysing persuasive technologies for health and wellness. In this study, these designs were found in social networking, shopping, leisure and reading apps which seek more clicks and monetisation from users. The persuasive designs in these apps do not contribute to user well-being and are misaligned with user interests. 

Persuasive triggers and reminders play crucial roles in cultivating users' habitual phone-checking behaviours. Both Fogg's Persuasive Design behaviour model \citep{fogg2009behavior} and the PSD model emphasised the role of triggers/reminders in strengthening users to perform target behaviours. In surveys and interviews, participants reported that a set of apps fostered their daily check-in habits. This habitual behaviour was triggered by phone vibrations, rings, reminders and flashing, which eventually led to users unlocking their phones unconsciously every 15-30 minutes, even without such triggers. Habitual checking is one of the symptoms of digital addiction and takes up a considerable amount of users’ screen time daily \citep{almourad2020defining,heitmayer2021smartphones}.

Some most frequently used PSD strategies, such as personalisation, reduction and rewards, might deprive users of the opportunity to make independent decisions in long-term use. Before video platforms and shopping apps introduced algorithmic recommendations, users had more time to autonomously explore different topics and products. Algorithmic recommendations have come to increasingly influence users' decision-making in many cases under the banner of convenience for users, but many interviewees in our study indicated that some smartphone apps powered by such algorithms ``know'' users to an uncomfortable degree. Ten survey participants complained that the videos recommended by TikTok were so addictive that they wasted ``too much time'' and ``lost control'' (see Table 2).

Besides popular persuasive design strategies, we found that some companies use manipulative and deceptive strategies to ``persuade'' users. One example is the WeChat display FOMO tags on top of the recommended articles to persuade users to spend more time on its services. On the one hand, \cite{li2022network} found that FOMO is positively associated with smartphone addiction. This repeated occurrence of the FOMO tag might increase users' level of FOMO, higher levels of FOMO were associated with a greater impact of social media on one's daily activities and productivity, particularly on messenger and social network use disorders \citep{rozgonjuk2020fear,przybylski2013motivational}. On the other hand, people who use their smartphones frequently for social purposes form smartphone habits more quickly, which might lead to PSU \citep{van2015modeling}. Another example is Pinduoduo, which applies gambling-like designs to exploit users' time and social contacts. Gambling disorder has been recognised as a behavioural addiction in DSM-5 \citep{edition2013diagnostic}. Deceiving users with nontransparent rules, decoy rewards, and interactive algorithms can lead to distress and self-blame among users, accustom them to gambling-like behaviours, and possibly become addicted to Pinduoduo.

Based on the above empirical evidence, we believe that persuasive designs influence the occurrence of PSU. When analysing PSU behaviours, it is crucial to consider persuasive designs. Interaction designers relying on such technologies need to consider the long-term impact of their products in terms of time consumption, habit cultivation, decision-making deprivation, behavioural addiction and human-computer relationships.

\subsection{Ethical implications for interaction designers}

Persuasion in interactive computing systems is becoming increasingly intelligent, subtle and influential. As Fogg pointed out, with interactive technologies, users receive information and respond immediately. Such interactive behaviour is different from traditional media (for example, TV and newspapers), which do not sustain interactive looping behaviours. Persuasive technology has amplified such interactive looping behaviours radically \citep{fogg2009behavior}. Persuasive technologies track and produce personalised reminders and suggestions on smartphones autonomously, instantly and persistently, and they are getting better at it every day.

We put more emphasis on investigating the relationship between persuasive designs and PSU in our study; however, we documented that nearly all participants also expressed positive sentiments towards the built-in features of smartphones, learning, hobby and time management apps. Participants believed these apps improved their quality of life and well-being, which were the initial goals of persuasive technology pioneers. According to the user screen time we have collected, users spend far less time using these persuasive technologies designed for user well-being, and most of their time is occupied by persuasive designs, such as for monetisation, that compromise their well-being. 

There are no governmental or industry regulations on persuasive technologies to block deception and manipulation. The authors observed some apps applying persuasive designs to exploit users: first, the abusive applications of tailoring and suggestions; for example, some trending articles on Zhihu and Weibo were actually paid promotions which target specific groups of users. Second, the overuse of reminders to seek users’ attention and exploit users' time, for example, the broad adoption of red dots and notifications on app icons. Third, some algorithms take advantage of users' weaknesses; for example, the addictive algorithm of TikTok has troubled many study participants. Troubled by these persuasive features, nonetheless, users cannot turn off these persuasive functions.

There are studies on how to design persuasive technologies ethically, approaches such as stakeholder analysis \citep{fogg2002persuasive}, moral principles \citep{berdichevsky1999toward}, voluntariness assessment \citep{smids2012voluntariness} and interdisciplinary research methods \citep{borgefalk2019ethics} were proposed by academia. Seven out of ten interviewees in our study have not heard of persuasive technology before the interviews, which might point toward a general lack of awareness of persuasive technology. Persuasive designs are often the default setting in operating systems and app installations, as far as we have observed in this study. The authors argue that one urgent ethical challenge interaction designers face is that most users are persuaded without their consent. Our users found themselves, not being persuaded with explicit arguments or reasons, but through the constant harvest of their private data and exploiting users' emotions. We do not know how to think of such means but as manipulative. In other words, the sort of means that Fogg warned against in his inceptional and seminal work on persuasive technology. Constant exposure to persuasive designs might lead to an exhaustion of self-control, which might be addressed by future work. 

\subsection{Limitations and future work} 

Our study has some limitations and open questions for future work remain. First, we used convenience sampling method, and the participants we reached were relatively small in number. The survey collected 183 valid results from Chinese university students; it is a small sample compared with the target group population size of 32.86 million \citep{Statista}. More questionnaires need to be distributed to study the group with improved sampling methods. Second, the self-reported screen time of the participants in this study differs from their actual screen time, as evidenced by the actual screen times (mean value=7.1, source: screenshots of their phones) of ten interviewees, which are longer than their self-reported screen times (mean value=6.0, source: questionnaires). Future studies need to examine the validity of self-reported measures of smartphone use and develop improved tools for quantifying media use \citep{parry2021systematic}. Screen time tracking apps could improve the accuracy of capturing usage, but it entails potential invasions of user privacy. We only used three indicators, i.e., loss of control, perceived negativeness and overuse, to identify potential PSU, Therefore, the findings of our study should be interpreted with caution and further research is needed to validate the scales used in our questionnaire. Third, our follow-up study to assess whether knowledge of persuasive technologies would change users' screen time is limited by participants having varying levels of knowledge on how to manage their screen time, since we did not standardise the information provided to them in case of questions at the end of the interviews. We are also unable to make statements about the reasons why participants changed screen time. Contextual factors might have played a role, for instance, sickness, family visits or holidays. Further research is needed to identify effective interventions for reducing PSU and improving digital well-being \citep{loid2020pop,olson2022nudge}. Fourth, we observed that the interviewees who had heard about persuasive technology before the interviews identified more persuasive designs than others. Even with the discussion of the definition and examples, the interviewees might need more time to understand and observe the persuasive designs from their smartphones.  

We recommend that future research investigates the impact of persuasive designs longitudinally. Currently, there is relatively little research on the abusive application of persuasive technology in commonly used apps and operating systems, which consumes much of young adults' screen time; there might also be other negative effects besides the ones we discussed.

\section{Conclusion}

We found that persuasive designs were perceived to prolong the screen time of the participants in our study and contributed to PSU. Participants reported that short video, social networking, gaming and learning apps most affected their attitude and behaviour, and these apps were found to employ multiple persuasive designs and occupy much of participants' screen time. The most frequently identified persuasive design strategies were reminders, personalisation, reduction, reward, suggestion and emotional motivators. These could have negative long-term impacts on users in relation to prolonging their screen time, reinforcing phone-checking habits and oversimplifying decision-making. Some ethically controversial strategies (persuasion without consent, abusively applying emotional triggers and using gambling-like designs) have been documented in our study. To answer the question we set out to answer – do persuasive designs make smartphones more addictive – we find indications that persuasive designs might contribute to problematic and addiction-related behaviours. We recommend HCI researchers and designers, as well as psychologists, examine the long-term impact of persuasive designs and other similar designs on their users.  


\begin{thebibliography}{64}
\expandafter\ifx\csname natexlab\endcsname\relax\def\natexlab#1{#1}\fi
\providecommand{\url}[1]{\texttt{#1}}
\providecommand{\href}[2]{#2}
\providecommand{\path}[1]{#1}
\providecommand{\DOIprefix}{doi:}
\providecommand{\ArXivprefix}{arXiv:}
\providecommand{\URLprefix}{URL: }
\providecommand{\Pubmedprefix}{pmid:}
\providecommand{\doi}[1]{\href{http://dx.doi.org/#1}{\path{#1}}}
\providecommand{\Pubmed}[1]{\href{pmid:#1}{\path{#1}}}
\providecommand{\bibinfo}[2]{#2}
\ifx\xfnm\relax \def\xfnm[#1]{\unskip,\space#1}\fi
\bibitem[{Adib and Orji(2021)}]{adib2021systematic}
\bibinfo{author}{Adib, A.}, \bibinfo{author}{Orji, R.}, \bibinfo{year}{2021}.
\newblock \bibinfo{title}{A systematic review of persuasive strategies in
  mobile e-commerce applications and their implementations}, in:
  \bibinfo{booktitle}{International Conference on Persuasive Technology},
  \bibinfo{organization}{Springer}. pp. \bibinfo{pages}{217--230}.
\bibitem[{Almourad et~al.(2020)Almourad, McAlaney, Skinner, Pleya and
  Ali}]{almourad2020defining}
\bibinfo{author}{Almourad, M.B.}, \bibinfo{author}{McAlaney, J.},
  \bibinfo{author}{Skinner, T.}, \bibinfo{author}{Pleya, M.},
  \bibinfo{author}{Ali, R.}, \bibinfo{year}{2020}.
\newblock \bibinfo{title}{Defining digital addiction: Key features from the
  literature}.
\newblock \bibinfo{journal}{Psihologija} \bibinfo{volume}{53},
  \bibinfo{pages}{237--253}.
\bibitem[{Berdichevsky and Neuenschwander(1999)}]{berdichevsky1999toward}
\bibinfo{author}{Berdichevsky, D.}, \bibinfo{author}{Neuenschwander, E.},
  \bibinfo{year}{1999}.
\newblock \bibinfo{title}{Toward an ethics of persuasive technology}.
\newblock \bibinfo{journal}{Communications of the ACM} \bibinfo{volume}{42},
  \bibinfo{pages}{51--58}.
\bibitem[{Billieux et~al.(2015a)Billieux, Maurage, Lopez-Fernandez, Kuss and
  Griffiths}]{billieux2015can}
\bibinfo{author}{Billieux, J.}, \bibinfo{author}{Maurage, P.},
  \bibinfo{author}{Lopez-Fernandez, O.}, \bibinfo{author}{Kuss, D.J.},
  \bibinfo{author}{Griffiths, M.D.}, \bibinfo{year}{2015}a.
\newblock \bibinfo{title}{Can disordered mobile phone use be considered a
  behavioral addiction? an update on current evidence and a comprehensive model
  for future research}.
\newblock \bibinfo{journal}{Current Addiction Reports} \bibinfo{volume}{2},
  \bibinfo{pages}{156--162}.
\bibitem[{Billieux et~al.(2015b)Billieux, Schimmenti, Khazaal, Maurage and
  Heeren}]{billieux2015we}
\bibinfo{author}{Billieux, J.}, \bibinfo{author}{Schimmenti, A.},
  \bibinfo{author}{Khazaal, Y.}, \bibinfo{author}{Maurage, P.},
  \bibinfo{author}{Heeren, A.}, \bibinfo{year}{2015}b.
\newblock \bibinfo{title}{Are we overpathologizing everyday life? a tenable
  blueprint for behavioral addiction research}.
\newblock \bibinfo{journal}{Journal of behavioral addictions}
  \bibinfo{volume}{4}, \bibinfo{pages}{119--123}.
\bibitem[{Blease(2015)}]{blease2015too}
\bibinfo{author}{Blease, C.}, \bibinfo{year}{2015}.
\newblock \bibinfo{title}{Too many ‘friends,’too few ‘likes’?
  evolutionary psychology and ‘facebook depression’}.
\newblock \bibinfo{journal}{Review of General Psychology} \bibinfo{volume}{19},
  \bibinfo{pages}{1--13}.
\bibitem[{Borgefalk and Leon(2019)}]{borgefalk2019ethics}
\bibinfo{author}{Borgefalk, G.}, \bibinfo{author}{Leon, N.d.},
  \bibinfo{year}{2019}.
\newblock \bibinfo{title}{The ethics of persuasive technologies in pervasive
  industry platforms: the need for a robust management and governance
  framework}, in: \bibinfo{booktitle}{International Conference on Persuasive
  Technology}, \bibinfo{organization}{Springer}. pp. \bibinfo{pages}{156--167}.
\bibitem[{Busch and McCarthy(2021)}]{busch2021antecedents}
\bibinfo{author}{Busch, P.A.}, \bibinfo{author}{McCarthy, S.},
  \bibinfo{year}{2021}.
\newblock \bibinfo{title}{Antecedents and consequences of problematic
  smartphone use: A systematic literature review of an emerging research area}.
\newblock \bibinfo{journal}{Computers in Human Behavior} \bibinfo{volume}{114},
  \bibinfo{pages}{106414}.
\bibitem[{Cellina et~al.(2021)Cellina, Marzetti and Gui}]{cellina2021self}
\bibinfo{author}{Cellina, F.}, \bibinfo{author}{Marzetti, G.V.},
  \bibinfo{author}{Gui, M.}, \bibinfo{year}{2021}.
\newblock \bibinfo{title}{Self-selection and attrition biases in app-based
  persuasive technologies for mobility behavior change: Evidence from a swiss
  case study}.
\newblock \bibinfo{journal}{Computers in Human Behavior} \bibinfo{volume}{125},
  \bibinfo{pages}{106970}.
\bibitem[{Cemiloglu et~al.(2021)Cemiloglu, Naiseh, Catania, Oinas-Kukkonen and
  Ali}]{cemiloglu2021fine}
\bibinfo{author}{Cemiloglu, D.}, \bibinfo{author}{Naiseh, M.},
  \bibinfo{author}{Catania, M.}, \bibinfo{author}{Oinas-Kukkonen, H.},
  \bibinfo{author}{Ali, R.}, \bibinfo{year}{2021}.
\newblock \bibinfo{title}{The fine line between persuasion and digital
  addiction}, in: \bibinfo{booktitle}{International Conference on Persuasive
  Technology}, \bibinfo{organization}{Springer}. pp. \bibinfo{pages}{289--307}.
\bibitem[{Chen et~al.(2017)Chen, Liu, Ding, Ying, Wang and
  Wen}]{chen2017gender}
\bibinfo{author}{Chen, B.}, \bibinfo{author}{Liu, F.}, \bibinfo{author}{Ding,
  S.}, \bibinfo{author}{Ying, X.}, \bibinfo{author}{Wang, L.},
  \bibinfo{author}{Wen, Y.}, \bibinfo{year}{2017}.
\newblock \bibinfo{title}{Gender differences in factors associated with
  smartphone addiction: a cross-sectional study among medical college
  students}.
\newblock \bibinfo{journal}{BMC psychiatry} \bibinfo{volume}{17},
  \bibinfo{pages}{1--9}.
\bibitem[{Clarke et~al.(2015)Clarke, Braun and Hayfield}]{clarke2015thematic}
\bibinfo{author}{Clarke, V.}, \bibinfo{author}{Braun, V.},
  \bibinfo{author}{Hayfield, N.}, \bibinfo{year}{2015}.
\newblock \bibinfo{title}{Thematic analysis}.
\newblock \bibinfo{journal}{Qualitative psychology: A practical guide to
  research methods} \bibinfo{volume}{222}, \bibinfo{pages}{248}.
\bibitem[{Daniel(2018)}]{daniel2018our}
\bibinfo{author}{Daniel, J.}, \bibinfo{year}{2018}.
\newblock \bibinfo{title}{Our letter to the apa}.
\newblock \URLprefix \url{https://screentimenetwork.org/apa}.
\bibitem[{Della~Vedova et~al.(2022)Della~Vedova, Covolo, Muscatelli, Loscalzo,
  Giannini and Gelatti}]{della2022psychological}
\bibinfo{author}{Della~Vedova, A.M.}, \bibinfo{author}{Covolo, L.},
  \bibinfo{author}{Muscatelli, M.}, \bibinfo{author}{Loscalzo, Y.},
  \bibinfo{author}{Giannini, M.}, \bibinfo{author}{Gelatti, U.},
  \bibinfo{year}{2022}.
\newblock \bibinfo{title}{Psychological distress and problematic smartphone
  use: two faces of the same coin? findings from a survey on young italian
  adults}.
\newblock \bibinfo{journal}{Computers in Human Behavior} \bibinfo{volume}{132},
  \bibinfo{pages}{107243}.
\bibitem[{Devincenzi et~al.(2017)Devincenzi, Kwecko, de~Toledo, Mota, Casarin
  and da~Costa~Botelho}]{devincenzi2017persuasive}
\bibinfo{author}{Devincenzi, S.}, \bibinfo{author}{Kwecko, V.},
  \bibinfo{author}{de~Toledo, F.P.}, \bibinfo{author}{Mota, F.P.},
  \bibinfo{author}{Casarin, J.}, \bibinfo{author}{da~Costa~Botelho, S.S.},
  \bibinfo{year}{2017}.
\newblock \bibinfo{title}{Persuasive technology: Applications in education},
  in: \bibinfo{booktitle}{2017 IEEE Frontiers in Education Conference (FIE)},
  \bibinfo{organization}{IEEE}. pp. \bibinfo{pages}{1--7}.
\bibitem[{Ding et~al.(2016)Ding, Xu, Chen and Xu}]{ding2016beyond}
\bibinfo{author}{Ding, X.}, \bibinfo{author}{Xu, J.}, \bibinfo{author}{Chen,
  G.}, \bibinfo{author}{Xu, C.}, \bibinfo{year}{2016}.
\newblock \bibinfo{title}{Beyond smartphone overuse: Identifying addictive
  mobile apps}, in: \bibinfo{booktitle}{Proceedings of the 2016 CHI Conference
  Extended Abstracts on Human Factors in Computing Systems}, pp.
  \bibinfo{pages}{2821--2828}.
\bibitem[{eCorner(2013)}]{Rosenstein}
\bibinfo{author}{eCorner, S.}, \bibinfo{year}{2013}.
\newblock \bibinfo{title}{Justin rosenstein: No dislike button on facebook.
  youtube.}
\newblock \URLprefix \url{https://www.youtube.com/watch?v=11WbGqALF_I}.
\bibitem[{Edition(2013)}]{edition2013diagnostic}
\bibinfo{author}{Edition, F.}, \bibinfo{year}{2013}.
\newblock \bibinfo{title}{Diagnostic and statistical manual of mental
  disorders}.
\newblock \bibinfo{journal}{Am Psychiatric Assoc} \bibinfo{volume}{21},
  \bibinfo{pages}{591--643}.
\bibitem[{Elhai et~al.(2016)Elhai, Levine, Dvorak and Hall}]{elhai2016fear}
\bibinfo{author}{Elhai, J.D.}, \bibinfo{author}{Levine, J.C.},
  \bibinfo{author}{Dvorak, R.D.}, \bibinfo{author}{Hall, B.J.},
  \bibinfo{year}{2016}.
\newblock \bibinfo{title}{Fear of missing out, need for touch, anxiety and
  depression are related to problematic smartphone use}.
\newblock \bibinfo{journal}{Computers in Human Behavior} \bibinfo{volume}{63},
  \bibinfo{pages}{509--516}.
\bibitem[{Elhai et~al.(2020a)Elhai, Yang, Fang, Bai and
  Hall}]{elhai2020depression}
\bibinfo{author}{Elhai, J.D.}, \bibinfo{author}{Yang, H.},
  \bibinfo{author}{Fang, J.}, \bibinfo{author}{Bai, X.}, \bibinfo{author}{Hall,
  B.J.}, \bibinfo{year}{2020}a.
\newblock \bibinfo{title}{Depression and anxiety symptoms are related to
  problematic smartphone use severity in chinese young adults: Fear of missing
  out as a mediator}.
\newblock \bibinfo{journal}{Addictive behaviors} \bibinfo{volume}{101},
  \bibinfo{pages}{105962}.
\bibitem[{Elhai et~al.(2020b)Elhai, Yang, McKay and Asmundson}]{elhai2020covid}
\bibinfo{author}{Elhai, J.D.}, \bibinfo{author}{Yang, H.},
  \bibinfo{author}{McKay, D.}, \bibinfo{author}{Asmundson, G.J.},
  \bibinfo{year}{2020}b.
\newblock \bibinfo{title}{Covid-19 anxiety symptoms associated with problematic
  smartphone use severity in chinese adults}.
\newblock \bibinfo{journal}{Journal of Affective Disorders}
  \bibinfo{volume}{274}, \bibinfo{pages}{576--582}.
\bibitem[{Eyal(2014)}]{eyal2014hooked}
\bibinfo{author}{Eyal, N.}, \bibinfo{year}{2014}.
\newblock \bibinfo{title}{Hooked: How to build habit-forming products}.
\newblock \bibinfo{publisher}{Penguin}.
\bibitem[{Fogg(1998)}]{fogg1998persuasive}
\bibinfo{author}{Fogg, B.J.}, \bibinfo{year}{1998}.
\newblock \bibinfo{title}{Persuasive computers: perspectives and research
  directions}, in: \bibinfo{booktitle}{Proceedings of the SIGCHI conference on
  Human factors in computing systems}, pp. \bibinfo{pages}{225--232}.
\bibitem[{Fogg(2002)}]{fogg2002persuasive}
\bibinfo{author}{Fogg, B.J.}, \bibinfo{year}{2002}.
\newblock \bibinfo{title}{Persuasive technology: using computers to change what
  we think and do}.
\newblock \bibinfo{journal}{Ubiquity} \bibinfo{volume}{2002},
  \bibinfo{pages}{2}.
\bibitem[{Fogg(2009)}]{fogg2009behavior}
\bibinfo{author}{Fogg, B.J.}, \bibinfo{year}{2009}.
\newblock \bibinfo{title}{A behavior model for persuasive design}, in:
  \bibinfo{booktitle}{Proceedings of the 4th international Conference on
  Persuasive Technology}, pp. \bibinfo{pages}{1--7}.
\bibitem[{Hamari et~al.(2014)Hamari, Koivisto and
  Pakkanen}]{hamari2014persuasive}
\bibinfo{author}{Hamari, J.}, \bibinfo{author}{Koivisto, J.},
  \bibinfo{author}{Pakkanen, T.}, \bibinfo{year}{2014}.
\newblock \bibinfo{title}{Do persuasive technologies persuade?-a review of
  empirical studies}, in: \bibinfo{booktitle}{International conference on
  persuasive technology}, \bibinfo{organization}{Springer}. pp.
  \bibinfo{pages}{118--136}.
\bibitem[{Heitmayer and Lahlou(2021)}]{heitmayer2021smartphones}
\bibinfo{author}{Heitmayer, M.}, \bibinfo{author}{Lahlou, S.},
  \bibinfo{year}{2021}.
\newblock \bibinfo{title}{Why are smartphones disruptive? an empirical study of
  smartphone use in real-life contexts}.
\newblock \bibinfo{journal}{Computers in Human Behavior} \bibinfo{volume}{116},
  \bibinfo{pages}{106637}.
\bibitem[{Horwood and Anglim(2019)}]{horwood2019problematic}
\bibinfo{author}{Horwood, S.}, \bibinfo{author}{Anglim, J.},
  \bibinfo{year}{2019}.
\newblock \bibinfo{title}{Problematic smartphone usage and subjective and
  psychological well-being}.
\newblock \bibinfo{journal}{Computers in Human Behavior} \bibinfo{volume}{97},
  \bibinfo{pages}{44--50}.
\bibitem[{Huang et~al.(2021)Huang, Lai, Xue, Zhang and Wang}]{huang2021network}
\bibinfo{author}{Huang, S.}, \bibinfo{author}{Lai, X.}, \bibinfo{author}{Xue,
  Y.}, \bibinfo{author}{Zhang, C.}, \bibinfo{author}{Wang, Y.},
  \bibinfo{year}{2021}.
\newblock \bibinfo{title}{A network analysis of problematic smartphone use
  symptoms in a student sample}.
\newblock \bibinfo{journal}{Journal of Behavioral Addictions}
  \bibinfo{volume}{9}, \bibinfo{pages}{1032--1043}.
\bibitem[{Kampik et~al.(2018)Kampik, Nieves and Lindgren}]{kampik2018coercion}
\bibinfo{author}{Kampik, T.}, \bibinfo{author}{Nieves, J.C.},
  \bibinfo{author}{Lindgren, H.}, \bibinfo{year}{2018}.
\newblock \bibinfo{title}{Coercion and deception in persuasive technologies},
  in: \bibinfo{booktitle}{20th International Trust Workshop (co-located with
  AAMAS/IJCAI/ECAI/ICML 2018), Stockholm, Sweden, 14 July, 2018},
  \bibinfo{organization}{CEUR-WS}. pp. \bibinfo{pages}{38--49}.
\bibitem[{Kaptein et~al.(2015)Kaptein, Markopoulos, De~Ruyter and
  Aarts}]{kaptein2015personalizing}
\bibinfo{author}{Kaptein, M.}, \bibinfo{author}{Markopoulos, P.},
  \bibinfo{author}{De~Ruyter, B.}, \bibinfo{author}{Aarts, E.},
  \bibinfo{year}{2015}.
\newblock \bibinfo{title}{Personalizing persuasive technologies: Explicit and
  implicit personalization using persuasion profiles}.
\newblock \bibinfo{journal}{International Journal of Human-Computer Studies}
  \bibinfo{volume}{77}, \bibinfo{pages}{38--51}.
\bibitem[{Kaptein et~al.(2010)Kaptein, Markopoulos, De~Ruyter and
  Aarts}]{kaptein2010persuasion}
\bibinfo{author}{Kaptein, M.C.}, \bibinfo{author}{Markopoulos, P.},
  \bibinfo{author}{De~Ruyter, B.}, \bibinfo{author}{Aarts, E.},
  \bibinfo{year}{2010}.
\newblock \bibinfo{title}{Persuasion in ambient intelligence}.
\newblock \bibinfo{journal}{Journal of Ambient Intelligence and Humanized
  Computing} \bibinfo{volume}{1}, \bibinfo{pages}{43--56}.
\bibitem[{Khalil and Abdallah(2013)}]{khalil2013harnessing}
\bibinfo{author}{Khalil, A.}, \bibinfo{author}{Abdallah, S.},
  \bibinfo{year}{2013}.
\newblock \bibinfo{title}{Harnessing social dynamics through persuasive
  technology to promote healthier lifestyle}.
\newblock \bibinfo{journal}{Computers in Human Behavior} \bibinfo{volume}{29},
  \bibinfo{pages}{2674--2681}.
\bibitem[{Kim et~al.(2015)Kim, Kim and Jee}]{kim2015relationship}
\bibinfo{author}{Kim, S.E.}, \bibinfo{author}{Kim, J.W.}, \bibinfo{author}{Jee,
  Y.S.}, \bibinfo{year}{2015}.
\newblock \bibinfo{title}{Relationship between smartphone addiction and
  physical activity in chinese international students in korea}.
\newblock \bibinfo{journal}{Journal of behavioral addictions}
  \bibinfo{volume}{4}, \bibinfo{pages}{200--205}.
\bibitem[{Kwon et~al.(2013a)Kwon, Kim, Cho and Yang}]{kwon2013smartphone}
\bibinfo{author}{Kwon, M.}, \bibinfo{author}{Kim, D.J.}, \bibinfo{author}{Cho,
  H.}, \bibinfo{author}{Yang, S.}, \bibinfo{year}{2013}a.
\newblock \bibinfo{title}{The smartphone addiction scale: development and
  validation of a short version for adolescents}.
\newblock \bibinfo{journal}{PloS one} \bibinfo{volume}{8},
  \bibinfo{pages}{e83558}.
\bibitem[{Kwon et~al.(2013b)Kwon, Lee, Won, Park, Min, Hahn, Gu, Choi and
  Kim}]{kwon2013development}
\bibinfo{author}{Kwon, M.}, \bibinfo{author}{Lee, J.Y.}, \bibinfo{author}{Won,
  W.Y.}, \bibinfo{author}{Park, J.W.}, \bibinfo{author}{Min, J.A.},
  \bibinfo{author}{Hahn, C.}, \bibinfo{author}{Gu, X.}, \bibinfo{author}{Choi,
  J.H.}, \bibinfo{author}{Kim, D.J.}, \bibinfo{year}{2013}b.
\newblock \bibinfo{title}{Development and validation of a smartphone addiction
  scale (sas)}.
\newblock \bibinfo{journal}{PloS one} \bibinfo{volume}{8},
  \bibinfo{pages}{e56936}.
\bibitem[{Lanette et~al.(2018)Lanette, Chua, Hayes and
  Mazmanian}]{lanette2018much}
\bibinfo{author}{Lanette, S.}, \bibinfo{author}{Chua, P.K.},
  \bibinfo{author}{Hayes, G.}, \bibinfo{author}{Mazmanian, M.},
  \bibinfo{year}{2018}.
\newblock \bibinfo{title}{How much is' too much'? the role of a smartphone
  addiction narrative in individuals' experience of use}.
\newblock \bibinfo{journal}{Proceedings of the ACM on Human-Computer
  Interaction} \bibinfo{volume}{2}, \bibinfo{pages}{1--22}.
\bibitem[{Lei et~al.(2020)Lei, Ismail, Mohammad and
  Yusoff}]{lei2020relationship}
\bibinfo{author}{Lei, L.Y.C.}, \bibinfo{author}{Ismail, M.A.A.},
  \bibinfo{author}{Mohammad, J.A.M.}, \bibinfo{author}{Yusoff, M.S.B.},
  \bibinfo{year}{2020}.
\newblock \bibinfo{title}{The relationship of smartphone addiction with
  psychological distress and neuroticism among university medical students}.
\newblock \bibinfo{journal}{BMC psychology} \bibinfo{volume}{8},
  \bibinfo{pages}{1--9}.
\bibitem[{Lewis(2017)}]{lewis2017our}
\bibinfo{author}{Lewis, P.}, \bibinfo{year}{2017}.
\newblock \bibinfo{title}{‘our minds can be hijacked’: the tech insiders
  who fear a smartphone dystopia}.
\newblock \bibinfo{journal}{The guardian} \bibinfo{volume}{6},
  \bibinfo{pages}{2017}.
\bibitem[{Li et~al.(2022)Li, Niu, Mei and Griffiths}]{li2022network}
\bibinfo{author}{Li, L.}, \bibinfo{author}{Niu, Z.}, \bibinfo{author}{Mei, S.},
  \bibinfo{author}{Griffiths, M.D.}, \bibinfo{year}{2022}.
\newblock \bibinfo{title}{A network analysis approach to the relationship
  between fear of missing out (fomo), smartphone addiction, and social
  networking site use among a sample of chinese university students}.
\newblock \bibinfo{journal}{Computers in Human Behavior} \bibinfo{volume}{128},
  \bibinfo{pages}{107086}.
\bibitem[{Loid et~al.(2020)Loid, T{\"a}ht and Rozgonjuk}]{loid2020pop}
\bibinfo{author}{Loid, K.}, \bibinfo{author}{T{\"a}ht, K.},
  \bibinfo{author}{Rozgonjuk, D.}, \bibinfo{year}{2020}.
\newblock \bibinfo{title}{Do pop-up notifications regarding smartphone use
  decrease screen time, phone checking behavior, and self-reported problematic
  smartphone use? evidence from a two-month experimental study}.
\newblock \bibinfo{journal}{Computers in Human Behavior} \bibinfo{volume}{102},
  \bibinfo{pages}{22--30}.
\bibitem[{Marciano et~al.(2021)Marciano, Schulz and
  Camerini}]{marciano2021smartphone}
\bibinfo{author}{Marciano, L.}, \bibinfo{author}{Schulz, P.J.},
  \bibinfo{author}{Camerini, A.L.}, \bibinfo{year}{2021}.
\newblock \bibinfo{title}{How smartphone use becomes problematic: Application
  of the alt-sr model to study the predicting role of personality traits}.
\newblock \bibinfo{journal}{Computers in Human Behavior} \bibinfo{volume}{119},
  \bibinfo{pages}{106731}.
\bibitem[{No{\"e} et~al.(2019)No{\"e}, Turner, Linden, Allen, Winkens and
  Whitaker}]{noe2019identifying}
\bibinfo{author}{No{\"e}, B.}, \bibinfo{author}{Turner, L.D.},
  \bibinfo{author}{Linden, D.E.}, \bibinfo{author}{Allen, S.M.},
  \bibinfo{author}{Winkens, B.}, \bibinfo{author}{Whitaker, R.M.},
  \bibinfo{year}{2019}.
\newblock \bibinfo{title}{Identifying indicators of smartphone addiction
  through user-app interaction}.
\newblock \bibinfo{journal}{Computers in human behavior} \bibinfo{volume}{99},
  \bibinfo{pages}{56--65}.
\bibitem[{Nystr{\"o}m and Stibe(2020)}]{nystrom2020persuasive}
\bibinfo{author}{Nystr{\"o}m, T.}, \bibinfo{author}{Stibe, A.},
  \bibinfo{year}{2020}.
\newblock \bibinfo{title}{When persuasive technology gets dark?}, in:
  \bibinfo{booktitle}{European, Mediterranean, and Middle Eastern Conference on
  Information Systems}, \bibinfo{organization}{Springer}. pp.
  \bibinfo{pages}{331--345}.
\bibitem[{Oinas-Kukkonen and Harjumaa(2009)}]{oinas2009persuasive}
\bibinfo{author}{Oinas-Kukkonen, H.}, \bibinfo{author}{Harjumaa, M.},
  \bibinfo{year}{2009}.
\newblock \bibinfo{title}{Persuasive systems design: Key issues, process model,
  and system features}.
\newblock \bibinfo{journal}{Communications of the association for Information
  Systems} \bibinfo{volume}{24}, \bibinfo{pages}{28}.
\bibitem[{Olson et~al.(2022a)Olson, Sandra, Chmoulevitch, Raz and
  Veissi{\`e}re}]{olson2022nudge}
\bibinfo{author}{Olson, J.A.}, \bibinfo{author}{Sandra, D.A.},
  \bibinfo{author}{Chmoulevitch, D.}, \bibinfo{author}{Raz, A.},
  \bibinfo{author}{Veissi{\`e}re, S.P.}, \bibinfo{year}{2022}a.
\newblock \bibinfo{title}{A nudge-based intervention to reduce problematic
  smartphone use: Randomised controlled trial}.
\newblock \bibinfo{journal}{International Journal of Mental Health and
  Addiction} , \bibinfo{pages}{1--23}.
\bibitem[{Olson et~al.(2022b)Olson, Sandra, Colucci, Al~Bikaii, Chmoulevitch,
  Nahas, Raz and Veissi{\`e}re}]{olson2022smartphone}
\bibinfo{author}{Olson, J.A.}, \bibinfo{author}{Sandra, D.A.},
  \bibinfo{author}{Colucci, {\'E}.S.}, \bibinfo{author}{Al~Bikaii, A.},
  \bibinfo{author}{Chmoulevitch, D.}, \bibinfo{author}{Nahas, J.},
  \bibinfo{author}{Raz, A.}, \bibinfo{author}{Veissi{\`e}re, S.P.},
  \bibinfo{year}{2022}b.
\newblock \bibinfo{title}{Smartphone addiction is increasing across the world:
  A meta-analysis of 24 countries}.
\newblock \bibinfo{journal}{Computers in Human Behavior} \bibinfo{volume}{129},
  \bibinfo{pages}{107138}.
\bibitem[{Orji and Moffatt(2018)}]{orji2018persuasive}
\bibinfo{author}{Orji, R.}, \bibinfo{author}{Moffatt, K.},
  \bibinfo{year}{2018}.
\newblock \bibinfo{title}{Persuasive technology for health and wellness:
  State-of-the-art and emerging trends}.
\newblock \bibinfo{journal}{Health informatics journal} \bibinfo{volume}{24},
  \bibinfo{pages}{66--91}.
\bibitem[{Oulasvirta et~al.(2012)Oulasvirta, Rattenbury, Ma and
  Raita}]{oulasvirta2012habits}
\bibinfo{author}{Oulasvirta, A.}, \bibinfo{author}{Rattenbury, T.},
  \bibinfo{author}{Ma, L.}, \bibinfo{author}{Raita, E.}, \bibinfo{year}{2012}.
\newblock \bibinfo{title}{Habits make smartphone use more pervasive}.
\newblock \bibinfo{journal}{Personal and Ubiquitous computing}
  \bibinfo{volume}{16}, \bibinfo{pages}{105--114}.
\bibitem[{Oyibo and Vassileva(2019)}]{oyibo2019relationship}
\bibinfo{author}{Oyibo, K.}, \bibinfo{author}{Vassileva, J.},
  \bibinfo{year}{2019}.
\newblock \bibinfo{title}{The relationship between personality traits and
  susceptibility to social influence}.
\newblock \bibinfo{journal}{Computers in Human Behavior} \bibinfo{volume}{98},
  \bibinfo{pages}{174--188}.
\bibitem[{Parry et~al.(2021)Parry, Davidson, Sewall, Fisher, Mieczkowski and
  Quintana}]{parry2021systematic}
\bibinfo{author}{Parry, D.A.}, \bibinfo{author}{Davidson, B.I.},
  \bibinfo{author}{Sewall, C.J.}, \bibinfo{author}{Fisher, J.T.},
  \bibinfo{author}{Mieczkowski, H.}, \bibinfo{author}{Quintana, D.S.},
  \bibinfo{year}{2021}.
\newblock \bibinfo{title}{A systematic review and meta-analysis of
  discrepancies between logged and self-reported digital media use}.
\newblock \bibinfo{journal}{Nature Human Behaviour} \bibinfo{volume}{5},
  \bibinfo{pages}{1535--1547}.
\bibitem[{Przybylski et~al.(2013)Przybylski, Murayama, DeHaan and
  Gladwell}]{przybylski2013motivational}
\bibinfo{author}{Przybylski, A.K.}, \bibinfo{author}{Murayama, K.},
  \bibinfo{author}{DeHaan, C.R.}, \bibinfo{author}{Gladwell, V.},
  \bibinfo{year}{2013}.
\newblock \bibinfo{title}{Motivational, emotional, and behavioral correlates of
  fear of missing out}.
\newblock \bibinfo{journal}{Computers in human behavior} \bibinfo{volume}{29},
  \bibinfo{pages}{1841--1848}.
\bibitem[{Rosenquist et~al.(2021)Rosenquist, Scott~Morton and
  Weinstein}]{rosenquist2021addictive}
\bibinfo{author}{Rosenquist, N.J.}, \bibinfo{author}{Scott~Morton, F.M.},
  \bibinfo{author}{Weinstein, S.}, \bibinfo{year}{2021}.
\newblock \bibinfo{title}{Addictive technology and its implications for
  antitrust enforcement}.
\newblock \bibinfo{journal}{Available at SSRN 3787822} .
\bibitem[{Rozgonjuk et~al.(2018)Rozgonjuk, Levine, Hall and
  Elhai}]{rozgonjuk2018association}
\bibinfo{author}{Rozgonjuk, D.}, \bibinfo{author}{Levine, J.C.},
  \bibinfo{author}{Hall, B.J.}, \bibinfo{author}{Elhai, J.D.},
  \bibinfo{year}{2018}.
\newblock \bibinfo{title}{The association between problematic smartphone use,
  depression and anxiety symptom severity, and objectively measured smartphone
  use over one week}.
\newblock \bibinfo{journal}{Computers in Human Behavior} \bibinfo{volume}{87},
  \bibinfo{pages}{10--17}.
\bibitem[{Rozgonjuk et~al.(2020a)Rozgonjuk, Sindermann, Elhai, Christensen and
  Montag}]{rozgonjuk2020associations}
\bibinfo{author}{Rozgonjuk, D.}, \bibinfo{author}{Sindermann, C.},
  \bibinfo{author}{Elhai, J.D.}, \bibinfo{author}{Christensen, A.P.},
  \bibinfo{author}{Montag, C.}, \bibinfo{year}{2020}a.
\newblock \bibinfo{title}{Associations between symptoms of problematic
  smartphone, facebook, whatsapp, and instagram use: An item-level exploratory
  graph analysis perspective}.
\newblock \bibinfo{journal}{Journal of Behavioral Addictions}
  \bibinfo{volume}{9}, \bibinfo{pages}{686--697}.
\bibitem[{Rozgonjuk et~al.(2020b)Rozgonjuk, Sindermann, Elhai and
  Montag}]{rozgonjuk2020fear}
\bibinfo{author}{Rozgonjuk, D.}, \bibinfo{author}{Sindermann, C.},
  \bibinfo{author}{Elhai, J.D.}, \bibinfo{author}{Montag, C.},
  \bibinfo{year}{2020}b.
\newblock \bibinfo{title}{Fear of missing out (fomo) and social media’s
  impact on daily-life and productivity at work: Do whatsapp, facebook,
  instagram, and snapchat use disorders mediate that association?}
\newblock \bibinfo{journal}{Addictive Behaviors} \bibinfo{volume}{110},
  \bibinfo{pages}{106487}.
\bibitem[{Rozgonjuk et~al.(2021)Rozgonjuk, Sindermann, Elhai and
  Montag}]{rozgonjuk2021comparing}
\bibinfo{author}{Rozgonjuk, D.}, \bibinfo{author}{Sindermann, C.},
  \bibinfo{author}{Elhai, J.D.}, \bibinfo{author}{Montag, C.},
  \bibinfo{year}{2021}.
\newblock \bibinfo{title}{Comparing smartphone, whatsapp, facebook, instagram,
  and snapchat: which platform elicits the greatest use disorder symptoms?}
\newblock \bibinfo{journal}{Cyberpsychology, behavior, and social networking}
  \bibinfo{volume}{24}, \bibinfo{pages}{129--134}.
\bibitem[{Smids(2012)}]{smids2012voluntariness}
\bibinfo{author}{Smids, J.}, \bibinfo{year}{2012}.
\newblock \bibinfo{title}{The voluntariness of persuasive technology}, in:
  \bibinfo{booktitle}{International Conference on Persuasive Technology},
  \bibinfo{organization}{Springer}. pp. \bibinfo{pages}{123--132}.
\bibitem[{Textor(2022)}]{Statista}
\bibinfo{author}{Textor, C.}, \bibinfo{year}{2022}.
\newblock \bibinfo{title}{Number of undergraduate students enrolled at public
  colleges and universities in china from 2010 to 2021}.
\newblock \URLprefix
  \url{https://www.statista.com/statistics/227028/number-of-students-at-universities-in-china/}.
\bibitem[{Timmer et~al.(2015)Timmer, Kool and Est}]{timmer2015ethical}
\bibinfo{author}{Timmer, J.}, \bibinfo{author}{Kool, L.}, \bibinfo{author}{Est,
  R.v.}, \bibinfo{year}{2015}.
\newblock \bibinfo{title}{Ethical challenges in emerging applications of
  persuasive technology}, in: \bibinfo{booktitle}{International Conference on
  Persuasive Technology}, \bibinfo{organization}{Springer}. pp.
  \bibinfo{pages}{196--201}.
\bibitem[{Van~Deursen et~al.(2015)Van~Deursen, Bolle, Hegner and
  Kommers}]{van2015modeling}
\bibinfo{author}{Van~Deursen, A.J.}, \bibinfo{author}{Bolle, C.L.},
  \bibinfo{author}{Hegner, S.M.}, \bibinfo{author}{Kommers, P.A.},
  \bibinfo{year}{2015}.
\newblock \bibinfo{title}{Modeling habitual and addictive smartphone behavior:
  The role of smartphone usage types, emotional intelligence, social stress,
  self-regulation, age, and gender}.
\newblock \bibinfo{journal}{Computers in human behavior} \bibinfo{volume}{45},
  \bibinfo{pages}{411--420}.
\bibitem[{Waskom(2021)}]{waskom2021seaborn}
\bibinfo{author}{Waskom, M.L.}, \bibinfo{year}{2021}.
\newblock \bibinfo{title}{Seaborn: statistical data visualization}.
\newblock \bibinfo{journal}{Journal of Open Source Software}
  \bibinfo{volume}{6}, \bibinfo{pages}{3021}.
\bibitem[{Williams(2018)}]{williams2018stand}
\bibinfo{author}{Williams, J.}, \bibinfo{year}{2018}.
\newblock \bibinfo{title}{Stand out of our light: freedom and resistance in the
  attention economy}.
\newblock \bibinfo{publisher}{Cambridge University Press}.
\bibitem[{Yang et~al.(2019)Yang, Asbury and Griffiths}]{yang2019chinese}
\bibinfo{author}{Yang, Z.}, \bibinfo{author}{Asbury, K.},
  \bibinfo{author}{Griffiths, M.D.}, \bibinfo{year}{2019}.
\newblock \bibinfo{title}{Do chinese and british university students use
  smartphones differently? a cross-cultural mixed methods study}.
\newblock \bibinfo{journal}{International Journal of Mental Health and
  Addiction} \bibinfo{volume}{17}, \bibinfo{pages}{644--657}.

\end{thebibliography}

\appendix
\section{Survey results}

The survey results (English translation version) can be downloaded from the following link: \\
http://doi.org/10.5281/zenodo.4934731

\pagebreak 

\section{Examples of persuasive designs}

\begin{figure}[ht]
     \centering
     \begin{subfigure}[b]{0.4\textwidth}
         \centering
         \includegraphics[width=\textwidth]{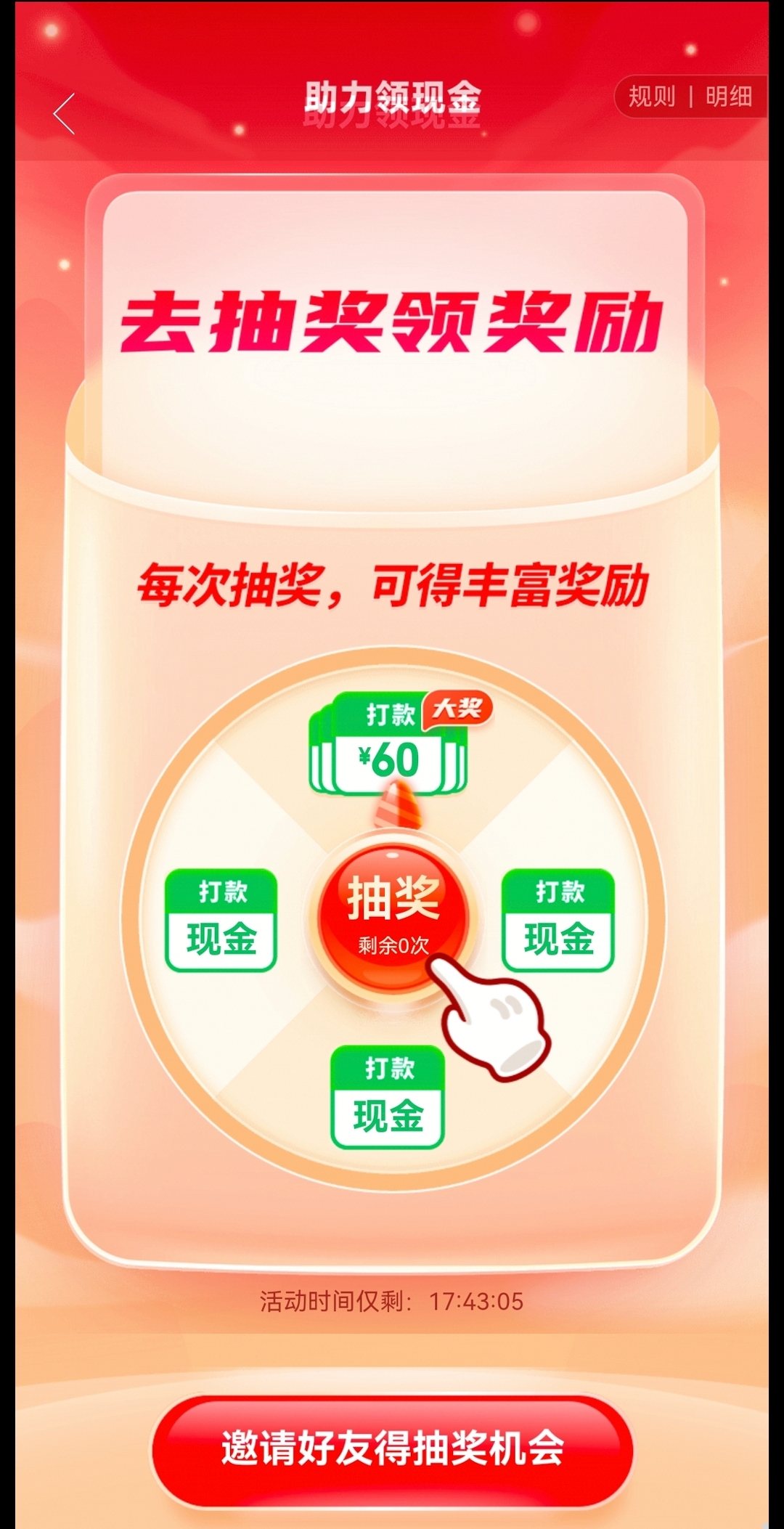}
         \caption{Gambling-like-design of Pinduoduo V6.57.0}
         \label{fig-4}
     \end{subfigure}
     \hfill
     \begin{subfigure}[b]{0.4\textwidth}
         \centering
         \includegraphics[width=\textwidth]{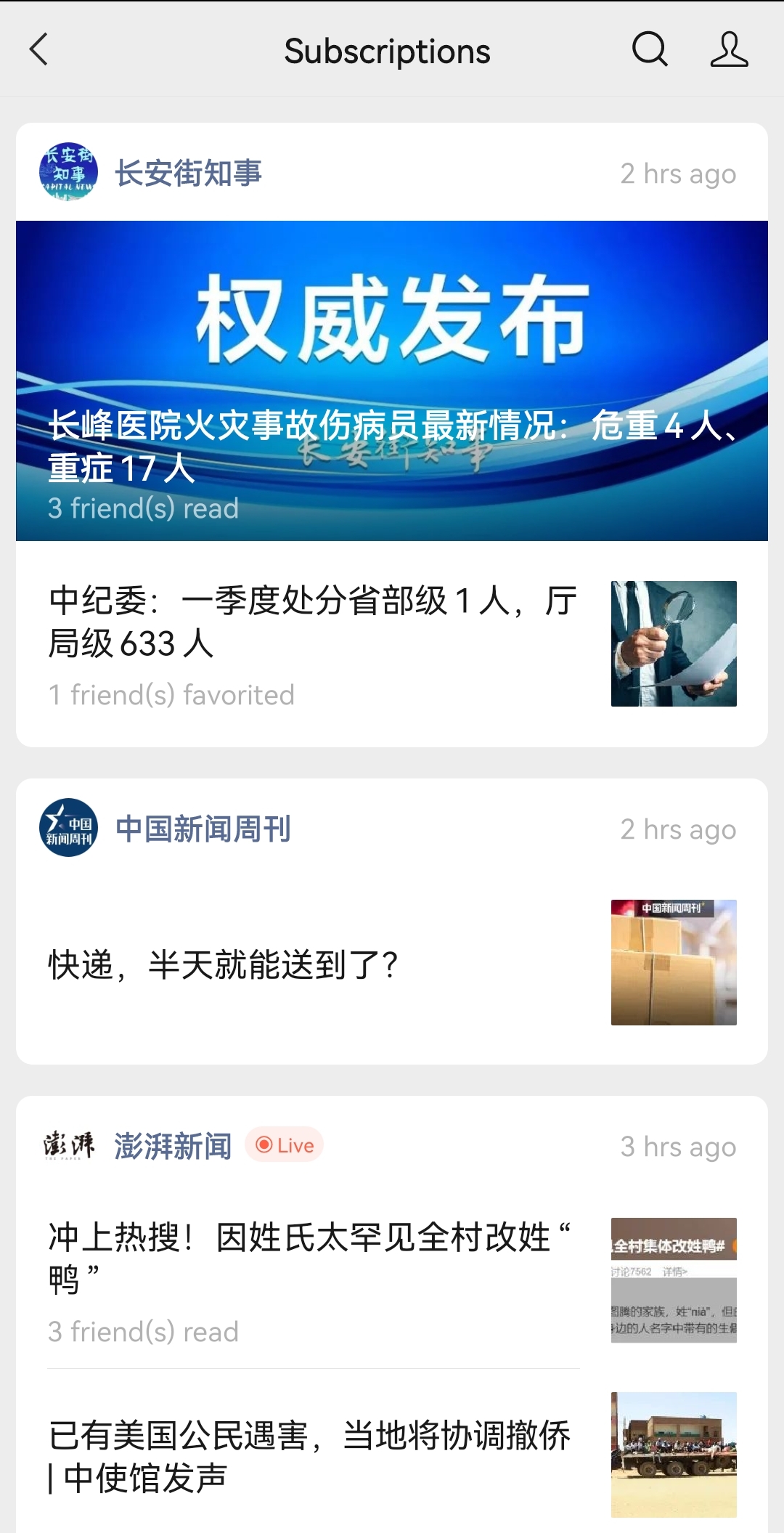}
         \caption{FOMO tags of WeChat V8.0.35}
         \label{fig-5}
     \end{subfigure}
     \hfill
     \caption{Screenshots of persuasive designs}
        \label{fig-screen}
\end{figure}

\end{document}